\documentclass[twocolumn]{aastex631}

\newcommand{\G}{\mathcal{G}}

\newcommand{\appropto}{\mathrel{\vcenter{\offinterlineskip\halign{\hfil$##$\cr\propto\cr\noalign{\kern2pt}\sim\cr\noalign{\kern-2pt}}}}}

\newcommand{\msun}{\ensuremath{\,{\rm M_\Sun}}}
\newcommand{\rsun}{\ensuremath{\,{\rm R_\Sun}}}

\newcommand{\mj}{\ensuremath{\,{\rm M_{\rm J}}}}
\newcommand{\rj}{\ensuremath{\,{\rm R_{\rm J}}}}
\newcommand{\degree}{\ensuremath{\,^{\circ}}}

\usepackage{url}
\usepackage{graphicx}
\usepackage{amsmath,amstext}
\usepackage[T1]{fontenc}

\usepackage{amsmath}
\usepackage{float}
\usepackage{longtable}
\usepackage{rotating}
\graphicspath{{./}{Figures/}}

\shorttitle{A Tendency Toward Alignment in Single-Star Warm Jupiter Systems}
\shortauthors{Rice et al.}

\begin{document}
\title{A Tendency Toward Alignment in Single-Star Warm Jupiter Systems}

\author[0000-0002-7670-670X]{Malena Rice} %confirmed
\altaffiliation{51 Pegasi b Fellow}
%\altaffiliation{NSF Graduate Research Fellow}
\affiliation{Department of Physics and Kavli Institute for Astrophysics and Space Research, Massachusetts Institute of Technology, Cambridge, MA 02139, USA}
\affiliation{Department of Astronomy, Yale University, New Haven, CT 06511, USA}

\author[0000-0002-7846-6981]{Songhu Wang} %confirmed
\affiliation{Department of Astronomy, Indiana University, Bloomington, IN 47405, USA}

\author[0000-0002-0376-6365]{Xian-Yu Wang} %confirmed %xianyu_wang@nao.cas.cn 
\affiliation{National Astronomical Observatories, Chinese Academy of Sciences, Beijing 10010, China}
\affiliation{University of the Chinese Academy of Sciences, Beijing, 100049, China}

\author[0000-0001-7409-5688]{Guðmundur Stefánsson} %confirmed
\altaffiliation{Henry Norris Russell Fellow}
\affiliation{Princeton University, Department of Astrophysical Sciences, 4 Ivy Lane, Princeton, NJ 08540, USA}

\author[0000-0002-0531-1073]{Howard Isaacson} %confirmed
\affiliation{Department of Astronomy, University of California Berkeley, Berkeley CA 94720, USA}
\affiliation{Centre for Astrophysics, University of Southern Queensland, Toowoomba, QLD, Australia}

\author[0000-0001-8638-0320]{Andrew W. Howard} %confirmed
\affiliation{Department of Astronomy, California Institute of Technology, Pasadena, CA 91125, USA}

\author[0000-0002-9632-9382]{Sarah E. Logsdon} %confirmed %slogsdon@astro.ucla.edu
\affil{NSF's National Optical-Infrared Astronomy Research Laboratory, 950 N. Cherry Ave., Tucson, AZ 85719, USA}

\author[0000-0001-9580-4869]{Heidi Schweiker} %Heidi.schweiker@noirlab.edu; confirmed
\affil{NSF's National Optical-Infrared Astronomy Research Laboratory, 950 N. Cherry Ave., Tucson, AZ 85719, USA}

\author[0000-0002-8958-0683]{Fei Dai} %confirmed
\affiliation{Division of Geological and Planetary Sciences 1200 E California Blvd, Pasadena, CA 91125}

\author[0000-0002-4480-310X]{Casey Brinkman} %confirmed
\altaffiliation{NSF Graduate Research Fellow}
\affiliation{Institute for Astronomy, University of Hawai'i at Manoa, Honolulu, HI 96822, USA}

\author[0000-0002-8965-3969]{Steven Giacalone} %confirmed
\affil{Department of Astronomy, University of California Berkeley, Berkeley, CA 94720-3411, USA}

\author[0000-0002-5034-9476]{Rae Holcomb}
\affil{Department of Physics and Astronomy, University of California Irvine, Irvine, CA 92697, USA}

\correspondingauthor{Malena Rice}
\email{malena.rice@yale.edu}

\begin{abstract}
The distribution of spin-orbit angles for systems with wide-separation, tidally detached exoplanets offers a unique constraint on the prevalence of dynamically violent planetary evolution histories. Tidally detached planets provide a relatively unbiased view of the primordial stellar obliquity distribution, since they cannot tidally realign within the system lifetime. We present the third result from our Stellar Obliquities in Long-period Exoplanet Systems (SOLES) survey: a measurement of the Rossiter-McLaughlin effect across two transits of the tidally detached warm Jupiter TOI-1478 b with the WIYN/NEID and Keck/HIRES spectrographs, revealing a sky-projected spin-orbit angle $\lambda=6.2^{+5.9}_{-5.5}\degree$. Combining this new measurement with the full set of archival obliquity measurements, including two previous constraints from the SOLES survey, we demonstrate that, in single-star systems, tidally detached warm Jupiters are preferentially more aligned than closer-orbiting hot Jupiters. This finding has two key implications: (1) planets in single-star systems tend to form within aligned protoplanetary disks, and (2) warm Jupiters form more quiescently than hot Jupiters, which, in single-star systems, are likely perturbed into a misaligned state through planet-planet interactions in the post-disk-dispersal phase. We also find that lower-mass Saturns span a wide range of spin-orbit angles, suggesting a prevalence of planet-planet scattering and/or secular mechanisms in these systems.
\end{abstract}


\section{Introduction} 
\label{section:intro}

In contrast with the quiescent picture presented by the solar system, exoplanets have been discovered orbiting forwards, sideways, and potentially even backwards around their host stars. The degree of spin-orbit alignment of a system is quantified by its ``stellar obliquity,'' which is the angle between the system's net orbital angular momentum vector and the spin axis of the host star. The maximum stellar obliquity attained by a system during its lifetime provides a unique constraint on its evolutionary history, offering a window into that system's dynamical past.  

Stellar obliquities are commonly constrained using the Rossiter-McLaughlin effect \citep{rossiter1924detection, mclaughlin1924some}, which offers an avenue to measure the 2D, sky-projected spin-orbit angle of an occulting body. The Rossiter-McLaughlin effect describes the phenomenon in which an occulter (in exoplanet systems, a transiting planet) blocks out different red- or blue-shifted components of its host star's light as it occults the stellar disk, producing radial velocity (RV) shifts across the transit. 

Because the Rossiter-McLaughlin effect is observed across individual transit events, the vast majority of stellar obliquity measurements to date have been made in systems with massive, close-orbiting planets due to the relatively short, frequent, and deep transits observed for these systems (see \citet{albrecht2022stellar} for a review of existing constraints). However, planets orbiting at small orbital separation are heavily impacted by tidal interactions with the host star, which can erase the signatures of misalignments over time. 

By contrast, ``tidally detached'' planets at wider orbital separation remain relatively unaffected by interactions with the host star over the age of the Universe. As a result, tidally detached planets provide a clearer view of the maximum stellar obliquity obtained in each system. Tidally detached planets can also help to distinguish whether misalignments typically occur at the protoplanetary disk stage -- that is, whether all planetary systems span a similar range of misalignments due to formation in misaligned disks -- or whether misalignments typically occur in the post-disk-dispersal phase.

The Stellar Obliquities in Long-period Exoplanet Systems (SOLES) survey \citep{rice2021soles, wang2022aligned} is, for the first time, systematically accruing a set of stellar obliquity constraints for the tidally detached population of exoplanets. These constraints will help to delineate the true rate of dynamically violent phases during planetary systems' evolution. By distinguishing the maximum misalignments typically attained by planetary systems as a function of orbital properties and planet mass, the SOLES survey will help to delineate the range of formation channels producing different types of planetary systems. 

An immediate outcome of this effort is a new set of constraints on the origins of hot and warm Jupiters. In this work, we present the third result from the SOLES survey: a sky-projected stellar obliquity measurement for the tidally detached warm Jupiter TOI-1478 b. We present two observations of the Rossiter-McLaughlin effect across two transits of TOI-1478 b, taken with the NEID spectrograph \citep{schwab2016design} on the WIYN 3.5-meter telescope and the High Resolution Echelle Spectrometer \citep[HIRES;][]{vogt1994hires} on the 10-meter Keck I telescope.

TOI-1478 b is a confirmed, tidally detached ($a/R_*=18.54_{-0.6}^{+0.7}$) giant planet ($M_p=0.851^{+0.052}_{-0.047} M_J$) initially identified through an analysis of full-frame images obtained by the Transiting Exoplanet Survey Satellite \citep[TESS;][]{ricker2015tess}. TOI-1478 b lies on a circular or near-circular orbit ($e=0.024^{+0.032}_{-0.017}$), with orbital period $P=10.180249 \pm 0.000015$ days \citep{rodriguez2021tess}. The planet is hosted by a solar-mass ($M_*=0.946^{+0.059}_{-0.041}M_{\odot}$) star, TOI-1478 (TIC 409794137), with $V=10.81$, $T_{\mathrm{eff}}=5595\pm83$ K, and age $\tau = 9.2^{+3.1}_{-3.9}$ Gyr \citep{rodriguez2021tess}.

Using our new observations, we find that TOI-1478 b is consistent with alignment, with $\lambda=6.2^{+5.9}_{-5.5}\degree$. Combining this finding with the previous SOLES results and archival measurements, we then demonstrate for the first time that warm Jupiters in single-star systems are preferentially more aligned than their hot Jupiter analogues. We conclude with implications of this finding, which suggests that (1) planetary systems around single stars are typically formed in aligned disks and (2) hot Jupiters are likely misaligned after disk dispersal, while warm Jupiters form quiescently.

\section{Observations}
\label{section:observations}
%We obtained radial velocity measurements with the Keck/HIRES instrument from 7:25-15:35 UT on Feb 22 for K2-261, and from 9:00-15:35 UT on Feb 24 for K2-140. Our data were reduced with the standard CPS pipeline. Conditions were good each night, with typical seeing ranging from $1.0-1.3\arcsec$. Humidity increased over the span of each observation, leading to degraded seeing towards egress and post-egress. For K2-140, a spike in humidity led to a dome closure towards the end of the observation, leading to a $\sim$40-minute gap in observations before the last radial velocity measurement.

\subsection{WIYN/NEID Observations} 
We obtained $20$ radial velocity measurements of TOI-1478 with the high-resolution mode of the NEID spectrograph ($R\sim110,000$) from 03:18-09:49 UT on Feb. 5, 2022. Exposure times were fixed at 1000 seconds. Observations were interrupted during the post-transit baseline from 08:39-08:57 UT to obtain a set of intermediate calibration frames, including a Fabry-P\'{e}rot etalon frame and three laser frequency comb frames. Excluding the intermediate calibration window, our observations spanned the full transit with an additional $\sim18$ and $\sim70$ minutes of pre- and post-transit observations, respectively, to constrain the RV baseline. Seeing was variable due to windy conditions, ranging from $1.0\arcsec-2.5\arcsec$. We provide NEID RV measurements and uncertainties in Table \ref{tab:rv_data_neid}.

To extract precise stellar parameters, we also obtained one 52-minute NEID exposure of TOI-1478 on March 6, 2022 (seeing ranged from $1.3\arcsec-1.9\arcsec$) and one 1-hour NEID exposure on March 12, 2022 (seeing stable at $\sim$1.0$\arcsec$). All data were reduced using the NEID Data Reduction Pipeline\footnote{More information on the NEID data reduction pipeline can be found here: https://neid.ipac.caltech.edu/docs/NEID-DRP/} and retrieved from the NExScI NEID Archive\footnote{https://neid.ipac.caltech.edu/}.

\begin{deluxetable}{ccccc}
\tablecaption{NEID radial velocities for the TOI-1478 system.\label{tab:rv_data_neid}}
\tabletypesize{\scriptsize}
\tablehead{
\colhead{Time (BJD)} & \colhead{RV (m/s)} & \colhead{$\sigma_{\rm RV}$ (m/s)}}
\tablewidth{300pt}
\startdata
2459615.655110 & 20799.5 & 5.0 \\
2459615.665986 & 20804.0 & 4.4 \\
2459615.678704 & 20803.8 & 4.6 \\
2459615.708742 & 20805.4 & 4.1 \\
2459615.734892 & 20792.6 & 4.2 \\
2459615.746750 & 20797.1 & 5.0 \\
2459615.759267 & 20795.1 & 5.2 \\
2459615.771037 & 20799.7 & 5.3 \\
2459615.782804 & 20788.8 & 4.9 \\
2459615.794706 & 20791.3 & 4.8 \\
2459615.806261 & 20793.2 & 5.1 \\
2459615.819525 & 20785.7 & 4.6 \\
2459615.830167 & 20787.1 & 4.1 \\
2459615.841928 & 20792.4 & 3.5 \\
2459615.853901 & 20795.9 & 4.0 \\
2459615.866358 & 20789.7 & 3.4 \\
2459615.890445 & 20788.7 & 3.0 \\
2459615.901364 & 20790.7 & 3.9 \\
2459615.915296 & 20790.2 & 4.2 \\
2459615.927363 & 20783.5 & 3.5\\
\enddata
\end{deluxetable}

\subsection{Keck/HIRES Observations}
We obtained 39 radial velocity measurements of TOI-1478 with the HIRES spectrograph from 06:00-13:00 UT on Feb. 15, 2022. The median exposure time was $592$ seconds, with 100k exposure meter counts per spectrum. Our observations spanned the full 4-hour transit of TOI-1478 b, with an additional $\sim$100 and $\sim$45 minutes of pre- and post-transit observations, respectively, to constrain the RV baseline. Seeing was stable throughout the observing period, ranging from $1.2\arcsec-1.4\arcsec$.

The moon phase was two days from full during our observation, with a separation of $36.6\degr-36.7\degr$ from TOI-1478 throughout the night. We used the C2 decker ($14\arcsec\times0.861\arcsec, R=60,000$) for all Keck/HIRES RV observations, enabling direct sky subtraction to better account for scattered light from the moon. Our dataset was reduced using the California Planet Search pipeline \citep{howard2010california}. We provide HIRES RV measurements, S-index values, and associated uncertainties in Table \ref{tab:rv_data_hires}.

To calibrate our observations and extract precise stellar parameters, we also obtained a 21-minute iodine-free HIRES template exposure of TOI-1478 using the B3 decker ($14.0\arcsec\times0.574\arcsec, R=72,000$) on UT Feb. 21, 2022. We collected 250k exposure meter counts for the template spectrum, which was observed in good conditions with $1.2\arcsec$ seeing.

\begin{deluxetable}{ccccc}
\tablecaption{HIRES radial velocities for the TOI-1478 system.\label{tab:rv_data_hires}}
\tabletypesize{\scriptsize}
\tablehead{
\colhead{Time (BJD)} & \colhead{RV (m/s)} & \colhead{$\sigma_{\rm RV}$ (m/s)} & \colhead{S-index} & \colhead{$\sigma_S$}}
\tablewidth{300pt}
\startdata
2459625.763126 & 7.29 & 1.37 & 0.153 & 0.001 \\
2459625.770626 & 9.27 & 1.38 & 0.155 & 0.001 \\
2459625.777964 & 4.25 & 1.32 & 0.155 & 0.001 \\
2459625.785418 & 8.66 & 1.44 & 0.155 & 0.001 \\
2459625.792536 & 1.63 & 1.36 & 0.155 & 0.001 \\
2459625.799538 & 6.65 & 1.34 & 0.153 & 0.001 \\
2459625.806147 & 8.54 & 1.31 & 0.152 & 0.001 \\
2459625.812859 & 6.17 & 1.34 & 0.153 & 0.001 \\
2459625.819630 & 7.46 & 1.28 & 0.153 & 0.001 \\
2459625.826262 & 6.78 & 1.32 & 0.154 & 0.001 \\
2459625.832894 & 3.34 & 1.31 & 0.155 & 0.001 \\
2459625.839537 & 3.57 & 1.31 & 0.153 & 0.001 \\
2459625.846632 & -0.22 & 1.47 & 0.153 & 0.001 \\
2459625.854098 & 3.47 & 1.39 & 0.152 & 0.001 \\
2459625.861366 & 4.48 & 1.26 & 0.151 & 0.001 \\
2459625.868333 & 8.35 & 1.38 & 0.151 & 0.001 \\
2459625.875300 & 6.66 & 1.30 & 0.151 & 0.001 \\
2459625.881851 & 10.19 & 1.39 & 0.152 & 0.001 \\
2459625.888981 & 1.70 & 1.33 & 0.150 & 0.001 \\
2459625.896273 & 2.31 & 1.32 & 0.152 & 0.001 \\
2459625.903645 & 4.20 & 1.36 & 0.151 & 0.001 \\
2459625.911388 & 2.25 & 1.42 & 0.151 & 0.001 \\
2459625.918667 & 2.11 & 1.37 & 0.153 & 0.001 \\
2459625.926156 & 1.24 & 1.44 & 0.151 & 0.001 \\
2459625.933066 & -2.40 & 1.50 & 0.152 & 0.001 \\
2459625.940647 & -2.17 & 1.45 & 0.150 & 0.001 \\
2459625.948228 & -4.86 & 1.29 & 0.152 & 0.001 \\
2459625.955716 & -10.76 & 1.26 & 0.150 & 0.001 \\
2459625.970819 & -6.96 & 1.32 & 0.151 & 0.001 \\
2459625.963181 & -8.71 & 1.22 & 0.151 & 0.001 \\
2459625.978389 & -15.52 & 1.26 & 0.149 & 0.001 \\
2459625.985738 & -12.07& 1.33 & 0.148 & 0.001 \\
2459625.993365 & -13.67 & 1.34 & 0.149 & 0.001 \\
2459626.001456 & -10.52 & 1.42 & 0.149 & 0.001 \\
2459626.010009 & -3.19 & 1.45 & 0.149 & 0.001 \\
2459626.018040 & -7.20 & 1.46 & 0.149 & 0.001 \\
2459626.026131 & -3.47 & 1.51 & 0.149 & 0.001 \\
2459626.035286 & -9.51 & 1.43 & 0.149 & 0.001 \\
2459626.044105 & -7.98 & 1.44 & 0.148 & 0.001 \\
\enddata
\end{deluxetable}

\section{Stellar Obliquity Modeling} 
\label{section:spinorbitmodel}

We applied the \texttt{allesfitter} Python package \citep{gunther2020allesfitter} to jointly model our in-transit Rossiter-McLaughlin data together with the photometric and radial velocity datasets presented in the TOI-1478 b discovery paper \citep{rodriguez2021tess}. These additional datasets include photometry from TESS, PEST, and FLWO/KeplerCam, as well as radial velocities from the TRES \citep{Furesz2008}, CHIRON \citep{Tokovinin2013}, FEROS \citep{Kaufer1999}, and CORALIE \citep{Queloz2001} spectrographs.

Our model parameters, each initialized with uniform priors provided in Table \ref{tab:system_properties}, include the orbital period ($P$), transit mid-times ($T_0$), cosine of the orbital inclination ($\cos{i}$), planet-to-star radius ratio ($R_{P}/R_{\star}$), sum of radii divided by the orbital semi-major axis ($(R_{\star}+R_{P})/a$), RV semi-amplitude ($K$), parameterized eccentricity and argument of periastron ($\sqrt{e}\,\cos{\,\omega}$, $\sqrt{e}\,\sin{\,\omega}$), sky-projected spin-orbit angle ($\lambda$), and sky-projected stellar rotational velocity ($v\sin i_{\star}$). The two limb darkening coefficients were initialized as $q_1=q_2=0.5$.

Initial guesses for $P$, $T_{0}$, $\cos{i}$, $R_{p}/R_{\star}$, $(R_{\star}+R_{p})/a$, $K$, $\sqrt{e}\,\cos{\,\omega}$, and $\sqrt{e}\,\sin{\,\omega}$ were drawn from \citet{rodriguez2021tess}. We applied a quadratic model to fit for additive RV offsets between each spectrograph, and we allowed $\lambda$ to vary between $-180\degr$ and $+180\degr$.

To thoroughly sample the posterior distribution for each model parameter, we ran an affine-invariant Markov Chain Monte Carlo (MCMC) analysis with 100 walkers. We ensured that each chain was fully converged by running all Markov chains to over 30$\times$ their autocorrelation length, which corresponded to at minimum 200,000 accepted steps per walker.

\begin{figure*}
    \centering
    \includegraphics[width=0.98\textwidth]{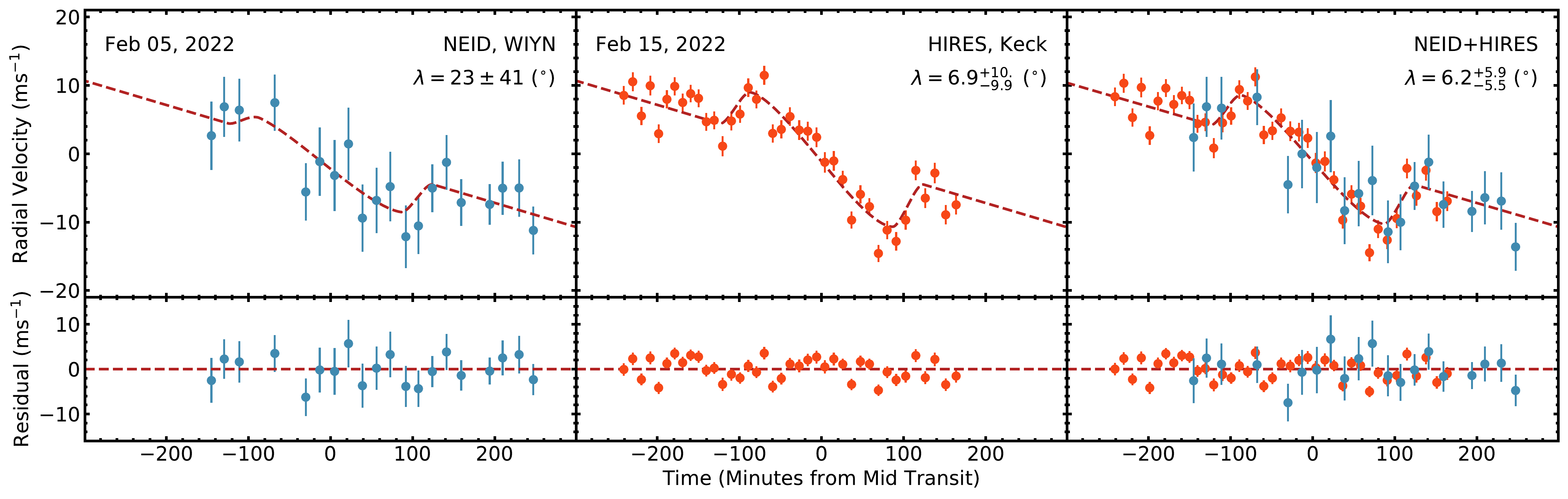}
    \caption{Best-fitting models for the Rossiter-McLaughlin observations obtained in this work. \textit{Left}: Best fit to the WIYN/NEID observations taken from the TOI-1478 b transit on UT Feb. 5, 2022. \textit{Center:} Best fit to the Keck/HIRES observations taken from the TOI-1478 b transit on UT Feb. 15, 2022. \textit{Right}: Joint fit incorporating both datasets.}
    \label{fig:rv_joint_fit}
\end{figure*}

The best-fit model parameters are provided in Table \ref{tab:system_properties} together with their associated 1$\sigma$ uncertainties. The resulting best-fit Rossiter-McLaughlin model from each of our three runs is shown in Figure \ref{fig:rv_joint_fit}, with the residuals provided below each model fit. We ultimately find that TOI-1478 is an aligned system, with sky-projected stellar obliquity $\lambda=6.2_{-5.5}^{+5.9}\degree$. An analysis of photometry from the TESS, WASP \citep{pollacco2006}, and KELT \citep{pepper2007kilodegree, pepper2012kelt, pepper2018kelt} surveys has revealed no periodicities that are robustly associated with the stellar spin rate \citep{rodriguez2021tess}, preventing a further constraint on the 3D stellar obliquity $\psi$.

\begin{deluxetable*}{lcccccc}
\tablecaption{System Parameters, Priors, and Results for TOI-1478 b  \label{tab:system_properties}}
\tabletypesize{\scriptsize}
\tablehead{
  \colhead{ } & \colhead{HIRES Spectrum} &\colhead{NEID Spectrum} & \colhead{Rodriguez+ 2021} &\colhead{Global RM fit: NEID+HIRES}      \\
%   \vspace{-1cm}
 \colhead{} & \colhead{The Cannon}  & \colhead{iSpec and PARAM 1.3}  & \colhead{EXOFASTv2}&\colhead{(Preferred Solution)}
}
\tablewidth{300pt}
\startdata
\multicolumn{5}{l}{Stellar Parameters:}\\
~~~~$M_*$ (\msun)   					& - 									& $0.907\pm0.022$  		    & $0.947^{+0.059}_{-0.041}$ & - \\
~~~~$R_*$ (\rsun) 						& -  									& $1.034\pm0.023$  			& $1.048^{+0.030}_{-0.029}$ & - \\
~~~~$\log{g}$ (cgs) 					& $4.231\pm0.164$ 	    	& $4.14\pm0.14$  			& $4.374^{+0.039}_{-0.032}$ & - \\
~~~~$[\rm{Fe/H}]$ (dex) 			    & $-0.03\pm0.06$ 			& $-0.17\pm0.07$ 		     	& $0.078^{+0.072}_{-0.065}$ & - \\
~~~~$T_{\rm eff}$ (K) 					& $5500\pm132$ 		     	& $5496\pm110$ 				& $5597^{+83}_{-82}$ & - \\
~~~~$v\sin i_{\star}$ (km/s) 		    & $3.00\pm0.98$ 			&$2.22\pm0.82$  				& $4.3\pm0.5^{a}$\ & $1.24\pm0.16$ \\
\\
\hline
\hline
  &Priors for global fit&Global fit 1: NEID&Global fit 2: HIRES &Global fit 3: NEID+HIRES  \\
  & & & &(Preferred Solution)  \\
\hline
\multicolumn{5}{l}{Fitted parameters:}\\
~~~~$R_P / R_\star$           &  $\mathcal U(0.1004; 0.0; 1.0)$    & $0.1021_{-0.0012}^{+0.0010}$ & $0.1020_{-0.0013}^{+0.0010}$ & $0.1020_{-0.0012}^{+0.0011}$   \\         
~~~~$(R_\star + R_p) / a$             &  $\mathcal U(0.04937; 0.0; 1.0)$  &   $0.0691_{-0.0038}^{+0.0042}$ & $0.0686_{-0.0040}^{+0.0044}$ & $0.0683\pm0.0032$            \\  
~~~~$\cos{i}$    &  $\mathcal U(0.026; 0.0; 1.0)$    & $0.0425_{-0.0049}^{+0.0058}$ & $0.0418_{-0.0053}^{+0.0060}$ & $0.0414\pm0.0044$     \\    
~~~~$T_{0} \ (\rm BJD_{\rm TDB} - 2450000) $     &    $\mathcal U(9066.014585;9065.0; 9067.0)$ 	& $9066.01871\pm0.00031$ & $9066.01870\pm0.00032$ & $9066.01870\pm0.00031$ &    \\  
~~~~$P$ (days)       &  $\mathcal U(10.180249; 9.0; 11.0)$             & $10.1802104\pm0.0000072$ & $10.1802116\pm0.0000071$ & $10.1802117\pm0.0000071$  \\ 
~~~~$K_{\rm}$ ($\rm m \ s^{-1}$)      &  $\mathcal U(82.5; 0.0; 1000.0)$    & $81.7\pm3.2$ & $81.8\pm3.2$ & $81.5\pm2.8$              \\  
~~~~$\sqrt{e} \cos{\omega}$         &  $\mathcal U(-0.053; -1.0; 1.0)$             & $-0.003_{-0.078}^{+0.073}$ & $-0.003\pm0.077$ & $-0.011\pm0.078$    \\        
~~~~$\sqrt{e} \sin{\omega}$         &  $\mathcal U(-0.145; -1.0; 1.0)$             & $0.236_{-0.16}^{+0.092}$ & $0.239_{-0.16}^{+0.093}$ & $0.222_{-0.11}^{+0.065}$     \\ 
~~~~$q_{\rm 1:HIRES}$     &  $\mathcal U(0.5;0;1)$    & $-$    &  $0.54\pm0.31$          &$0.49_{-0.30}^{+0.32}$        \\
~~~~$q_{\rm 2:HIRES}$     &  $\mathcal U(0.5;0;1)$    & $-$    &   $0.42_{-0.29}^{+0.36}$          &$0.39_{-0.27}^{+0.36}$        \\
~~~~$q_{\rm 1:NEID}$     &  $\mathcal U(0.5;0;1)$    & $0.41_{-0.29}^{+0.37}$    &  $-$          & $0.45_{-0.31}^{+0.35}$       \\
~~~~$q_{\rm 2:NEID}$     &  $\mathcal U(0.5;0;1)$    & $0.48_{-0.33}^{+0.35}$    &  $-$          &$0.52_{-0.36}^{+0.33}$        \\
~~~~$\lambda$ (deg)      &  $\mathcal U(0; -180; +180)$      &$23\pm41$ & $6.9_{-9.9}^{+10.}$ & $6.2_{-5.5}^{+5.9}$         \\ 
~~~~$v\sin i_{\star}$ ($\rm km \ s^{-1}$)          &  $\mathcal U(4.3; 0.0; 10)$     & $0.78_{-0.49}^{+0.53}$ & $1.31_{-0.19}^{+0.20}$ & $1.24\pm0.16$             \\  
\multicolumn{5}{l}{Derived parameters:}\\
~~~~$R_{P} \ (\rj)$        & -          &$1.07_{-0.10}^{+0.16}$ & $1.07_{-0.10}^{+0.16}$ & $1.07_{-0.10}^{+0.16}$     \\        
~~~~$M_{P} \ (\mj)$        & -          &$0.90\pm0.13$ & $0.89\pm0.13$ & $0.88_{-0.12}^{+0.11}$         \\ 
~~~~$b$            & -          & $0.641_{-0.034}^{+0.027}$ & $0.636_{-0.038}^{+0.029}$ & $0.636_{-0.035}^{+0.029}$         \\   
~~~~$T_{\rm 14}$            & -          & $4.140\pm0.042$ & $4.129\pm0.042$ & $4.131\pm0.041$        \\  
~~~~$\delta_{\rm FLWO}$            & -          & $11.52\pm0.18$ &  $11.51_{-0.16}^{+0.19}$ & $11.52\pm0.18$         \\  
~~~~$\delta_{\rm PESTR}$            & -          & $11.02_{-0.23}^{+0.28}$ & $11.02_{-0.24}^{+0.30}$ & $11.00_{-0.26}^{+0.28}$        \\  
~~~~$\delta_{\rm TESS}$            & -          & $10.95\pm0.10$ & $10.942\pm0.095$ & $10.936_{-0.087}^{+0.10}$        \\  
~~~~$a$            & -          & $0.0808_{-0.0087}^{+0.013}$ & $0.0814_{-0.0090}^{+0.013}$ & $0.0818_{-0.0085}^{+0.013}$         \\  
~~~~$i$ (deg)            & -          & $87.56_{-0.33}^{+0.28}$ & $87.60_{-0.35}^{+0.30}$ & $87.63\pm0.25$         \\            
~~~~$e$       & -          & $0.061_{-0.043}^{+0.050}$ & $0.062_{-0.044}^{+0.051}$ & $0.055\pm0.032$        \\     
~~~~$\omega$            & -          & $92_{-18}^{+35}$ & $92_{-18}^{+34}$ & $94_{-20.}^{+26}$          \\  
~~~~$u_{\rm 1:HIRES}$     &  -    & $-$    &  $0.54_{-0.38}^{+0.53}$          &$0.47_{-0.34}^{+0.50}$        \\
~~~~$u_{\rm 2:HIRES}$     &  -    & $-$    &  $0.10\pm0.44$        &$0.13\pm0.42$       \\
~~~~$u_{\rm 1:NEID}$     &  -    &$0.52_{-0.38}^{+0.58}$    &  $-$          &  $0.59_{-0.42}^{+0.57}$      \\
~~~~$u_{\rm 2:NEID}$     &  -    & $0.02\pm0.41$   &  $-$          &$-0.02_{-0.41}^{+0.43}$      \\
\enddata 
\tablenotetext{a}{\cite{rodriguez2021tess} measured $v\sin i_{\star}$ for TOI-1478 from TRES data using the methodology in \cite{zhou2018warm}.}
\end{deluxetable*}

\section{Stellar Parameters} 
\label{section:stellar_parameters}

Next, we analyzed individual iodine-free spectra of TOI-1478 to extract the system's stellar parameters directly from the HIRES and NEID observations described in this work. The observed RM semi-amplitude ($A_{\mathrm{RM}}$) was significantly smaller than anticipated based on previous constraints from \citet{rodriguez2021tess}: rather than detecting the anticipated signal with RM semi-amplitude $28.2\pm4.5$ m/s (obtained by simulating the system with the \texttt{allesfitter} Python package), we observed an RM semi-amplitude of only $6.6\pm0.9$ m/s. 

The theoretical semi-amplitude of a Rossiter-McLaughlin signal is given by

\begin{equation}
    A_{\mathrm{RM}} \simeq \frac{2}{3}\Big(\frac{R_p}{R_*}\Big)^2 v\sin i_* \sqrt{1-b^2},
\end{equation}
where $R_p$ and $R_*$ are the planet and stellar radius, respectively, $v\sin i_*$ is the projected stellar rotational velocity, and $b$ is the impact parameter of the planetary companion's transit \citep{triaud2017rossiter}. The transit depth $\frac{R_p}{R_*}$ and the impact parameter $b$ are each well-constrained by the observed transit, suggesting that an overestimated value of $v\sin i_*$ may have caused the observed discrepancy. As described in the remainder of Section \ref{section:stellar_parameters}, we obtain lower $v\sin i_*$ estimates from the HIRES and NEID data, which helps to account for the lower observed RM semi-amplitude.

\subsection{Stellar Parameters from Keck/HIRES}
\label{subsection:stellar_params_HIRES}
First, we modeled the stellar parameters of TOI-1478 by analyzing the iodine-free Keck/HIRES template spectrum obtained in this work. We applied the methods described in \citet{rice2020stellar}, which transfers stellar labels using the generative machine learning program \textit{The Cannon} \citep{ness2015cannon} after training on Keck/HIRES spectra of 1,202 FGK stars from the Spectral Properties of Cool Stars (SPOCS) catalogue \citep{valenti2005spectroscopic, brewer2016spectral}. The SPOCS catalogue includes 18 precisely determined stellar labels: 3 global stellar parameters ($T_{\mathrm{eff}}$, $\log g$, $v\sin i_{\star}$), and 15 elemental abundances: C, N, O, Na, Mg, Al, Si, Ca, Ti, V, Cr, Mn, Fe, Ni, and Y. 

We fit and divided out the continuum baseline of our Keck/HIRES template spectrum using the Alpha-shape Fitting to Spectrum (AFS) algorithm described in \citet{xu2019modeling}. Then, we calibrated the wavelength solution by cross-correlating the normalized spectrum with the solar atlas provided by \citet{wallace2011optical}. These two steps placed our spectrum on a normalized scale with a continuum baseline of unity for direct comparison with the training set.

To estimate the uncertainties in our results, we trained and applied our model separately using each of the 16 echelle orders in our Keck/HIRES spectrum. We then extracted the mean and standard deviation determined for each parameter. Our derived parameters and associated uncertainties are provided in Table \ref{tab:system_properties}. For direct comparison with previous results, we report only $\log g$, [Fe/H], $T_{\mathrm{eff}}$, and $v\sin i_*$ in this work. From our spectroscopic modeling of our Keck/HIRES spectrum, we obtained $v\sin i_* = 3.00\pm0.98$ km/s.

\subsection{Stellar Parameters from WIYN/NEID}
\label{subsection:stellar_params_NEID}
Next, we co-added our two NEID template spectra from Mar. 6 and 12, 2022, and we analyzed the combined spectrum to independently extract stellar parameters for TOI-1478. For this spectrum, we obtained a signal-to-noise ratio (SNR) of 58 for the wavelength range $480 - 680$ nm. We adopted the synthetic spectral fitting technique implemented in the \texttt{iSpec} Python package \citep{blanco2014determining, blanco2019modern} to determine the stellar parameters. 
%exposure time: 3109 s

%iSpec is a software framework that can apply the synthetic spectral fitting technique and the equivalent width method to derived atmosphere parameters. It offers a wide variety of radiative transfer codes, model atmospheres, solar abundances, and atomic line lists. 

Within \texttt{iSpec}, we selected the SPECTRUM radiative transfer code \citep{Gray1994} and the MARCS\footnote{\url{https://marcs.astro.uu.se/}} model atmospheres and solar abundances from \cite{Grevesse2007}. We adopted the sixth version of the GES atomic line list \citep{Heiter2015}. \texttt{iSpec} incorporates these constraints to iteratively generate and compare model spectra with the input spectrum, minimizing the difference using the nonlinear least-squares Levenberg-Marquardt fitting algorithm \citep{Markwardt2009}.

To reduce the computation time for this algorithm, we optimized specific spectral regions geared towards our parameters of interest. These include the wings of the H$\alpha$, H$\beta$, and Mg I triplet lines, which are sensitive to $T_{\mathrm{eff}}$ and $\log g$, as well as the Fe I and Fe II lines that can precisely constrain [Fe/H] and $v\sin i_*$. We used the derived $T_{\mathrm{eff}}$ and [Fe/H] values, as well as the $V$ magnitude ($V=10.81$) and corrected Gaia parallax \citep{Stassun2018} for TOI-1478, as inputs to the Bayesian stellar parameter estimation code \texttt{PARAM 1.3}\footnote{\url{http://stev.oapd.inaf.it/cgi-bin/param_1.3}} \citep{da2006basic} to extract the stellar mass ($M_*$) and radius ($R_*$). All of the parameters derived from this study are provided in Table \ref{tab:system_properties}. From this analysis, we obtained $v\sin i_* = 2.22\pm0.82$ km/s.

% \subsubsection{Stellar Parameters from MIST SED Fitting}
% We also derived stellar parameters by applying MESA Isochrones $\&$ Stellar Tracks \citep[MIST;][]{dotter2016mesa} stellar evolutionary models to fit the spectral energy distribution (SED) for TOI-1478.
% [Xian-Yu]

\subsection{Stellar Parameters from Global Fitting}
\label{subsection:stellar_params_global_fitting}
We also obtained an estimated $v\sin i_*$ value from our global fitting, which incorporated our full photometric and radial velocity datasets (see Section \ref{section:spinorbitmodel}). Based on this global model, we determined that $v\sin i_* = 1.24\pm0.16$ km/s. This lower $v\sin i_*$ value, which we adopt as our preferred solution, can fully account for the observed discrepancy between the expected and observed RM semi-amplitude.

\subsection{Comparison of Stellar Parameters}
\label{subsection:comparison_stellar_parameters}
All of our derived stellar parameters agree with previous results from \cite{rodriguez2021tess} at the 1$\sigma$ level, with the exception of [Fe/H] and $v\sin i_{\star}$. By comparison with the results from \cite{rodriguez2021tess}, we obtained a lower value of both [Fe/H] and $v\sin i_{\star}$ from the NEID spectrum. All results from the NEID spectrum are consistent with our results from the HIRES spectrum within 1$\sigma$.

Our derived values for $v\sin i_*$ each suggest a lower projected spin rate than the value reported in \citet{rodriguez2021tess}. This discrepancy in $v\sin i_*$ measurements likely results from the comparatively lower resolving power of the TRES spectrograph ($R=44,000$), by contrast with the NEID ($R=110,000$) and HIRES ($R=72,000$) spectrographs.

Measurements of $v\sin i_*$ are limited by the resolving power of the instrument: for example, previous work has demonstrated that the IGRINS spectrograph \citep[$R=45,000$;][]{park2014design}, with a comparable resolving power to that of TRES, has a lower measurement limit of $v\sin i_* \sim 3-4$ km/s \citep{kesseli2018magnetic, nofi2021projected}. As a result, the higher measured $v\sin i_*$ value in \citet{rodriguez2021tess} likely reflects the limited resolving power of TRES.

\section{Population Analysis}
\label{section:population_analysis} 
Results from the SOLES survey, together with archival measurements, are beginning to accrue into a statistical sample of stellar obliquities in tidally detached exoplanet systems. This enables, for the first time, comparative studies between distinct populations as the set of available constraints expands. 

Most stellar obliquity measurements to date provide constraints on only a 2D, sky-projected version of the true stellar obliquity. These measurements, including that of TOI-1478 b, are not fully representative of the true stellar obliquity for each individual system: systems that are aligned in their sky-projection may still be misaligned along the line of sight. However, population studies can provide useful insights into the typical trends across a group of planets, even while individual systems require further scrutiny to understand the typical degree of misalignment.

%We remove the warm Saturn WASP-49 b from our sample due to its large obliquity uncertainty ($\lambda=54^{+79}_-{58}\degree$). 

In this section, we examine the sky-projected stellar obliquity distribution in relatively isolated, single-star systems. We demonstrate that, in single-star systems, tidally detached Jovian planets are preferentially more aligned than their closer-orbiting counterparts. This result has two key implications that we discuss in this section: (1) protoplanetary disks are typically aligned in single-star systems, and (2) warm Jupiters typically form quiescently, while this is not necessarily true for hot Jupiters.

We emphasize that this trend toward alignment does not necessarily extend to tidally detached Jovian planets in binary and multi-star systems. Our choice to focus on only single-star systems excludes some highly misaligned warm Jupiters in systems with more than one star (e.g. K2-290 c \citep{hjorth2021backward}, Kepler-420 b \citep{santerne2014sophie}, and WASP-8 b \citep{queloz2010wasp, bourrier2017refined}). However, other warm Jupiters in binary or multi-star systems have instead been observed in or near alignment (e.g. V1298 Tau b \citep{gaidos2022zodiacal, johnson2022aligned} and WASP-11 b \citep{mancini2015gaps}). The set of obliquity constraints in binary star systems so far appears to be more heterogeneous than in the sample of single-star systems, and further measurements are required to delineate the dominant trends in this population.

\subsection{Sample Selection}
\label{subsection:sample_selection} 

\subsubsection{Data Sources}
Our sample consists of the set of systems included in both the NASA Exoplanet Archive ``Planetary Systems'' list \citep{ps_NEXScI} and the TEPCat catalogue \citep{southworth2011homogeneous}, which compiles published stellar obliquity measurements into a single database. Both datasets were downloaded on Feb. 19, 2022. We drew $\lambda$ and $T_{\mathrm{eff}}$ from the TEPCat catalogue, and, where possible, all other system parameters were drawn from the default parameter set of the NASA Exoplanet Archive. When parameters were unavailable in the NASA Exoplanet Archive, values were supplemented from the Extrasolar Values Encyclopaedia\footnote{\url{http://exoplanet.eu}} where possible.

We used the most recently obtained stellar obliquity measurement for most stars, with the exception of systems in which previous measurements were much more precise. To maximize the uniformity of our sample, we also favored measurements of the Rossiter-McLaughlin effect or Doppler tomography over constraints from starspot crossings. Where only upper planet mass and/or eccentricity limits were provided, we adopted the upper limit as the fiducial value. The full set of parameters used in this work is provided in Table \ref{tab:population_parameters}.

\subsubsection{Stellar Multiplicity}
To develop a clean sample of single-star systems, we then evaluated the stellar multiplicity of each system with a measured $\lambda$ value. We first cross-matched our sample with with the Catalogue of Exoplanets in Binary Star Systems \citep{schwarz2016new}, which includes binary and multiple-star systems with confirmed exoplanets. We removed all systems with companions listed either in this catalogue or elsewhere from a literature search of each system.

Next, we searched for any comoving companions in the SIMBAD database \citep{wenger2000simbad}, which includes parallaxes and proper motions drawn from Gaia eDR3 \citep{brown2021gaia}. We followed the methods of \citet{el2021million} to identify neighboring sources, with details summarized here for convenience.

We first queried SIMBAD for all stars with a projected separation $\theta$ within 5 pc of the host star, or 

\begin{equation}
    \theta\leq17.19\, \mathrm{arcmin} \times (\varpi\,\mathrm{mas}^{-1})
\end{equation}
for parallax $\varpi$. For each candidate companion star, we checked the following conditions to verify whether it is a comoving companion:

\vspace{3mm}
\noindent\textit{Condition 1:} The candidate companion must have a proper motion $\mu$ within 5 km/s of the planet-hosting star, or

\begin{equation}
\sqrt{\Delta \mu_{\mathrm{RA}}^2 + \Delta \mu_{\mathrm{Dec}}^2} \leq 1.05\, \mathrm{mas/yr} \times (\varpi_1 - 2\sigma_{\varpi_1})\, \mathrm{mas}^{-1}
\end{equation} 

\begin{equation}
    \Delta \mu_{\mathrm{RA}} = \mu_{\mathrm{RA}, 1} - \mu_{\mathrm{RA}, 2}
\end{equation}

\begin{equation}
    \Delta \mu_{\mathrm{Dec}} = \mu_{\mathrm{Dec}, 1} - \mu_{\mathrm{Dec}, 2}.
\end{equation}

\vspace{3mm}

\noindent\textit{Condition 2:} The candidate companion must have a parallax consistent with that of the planet-hosting star within 5$\sigma$, or 

\begin{equation}
|\varpi_1 - \varpi_2| \leq 5\sqrt{\sigma_{\varpi_1}^2 + \sigma_{\varpi_2}^2}.
\end{equation}

We note that \citet{el2021million} required a 2$\sigma$ agreement for Condition 2, whereas we set a more conservative 5$\sigma$ requirement to ensure a clean sample of single stars. We removed all systems from our sample in which one or more neighboring stars was identified as a comoving companion (that is, all systems in which at least one companion met both of the conditions above). 

Afterwards, we cross-validated our results with the the visual binary catalogue published by \citet{fontanive2021census}, which identified exoplanetary systems in binary or multiple-star configurations based on results from Gaia DR2 \citep{Gaia2016, gaia2018}. All of our results match with the classifications determined by \citet{fontanive2021census}. Table \ref{tab:population_parameters} includes the full set of single-star systems retained within our population study.

\subsection{Stellar Obliquities Comparison}
\label{subsection:obliquities_comparison} 
To compare the rate of misalignments across populations, we assigned a label of either ``aligned'' or ``misaligned'' to each system in our sample. We consider ``aligned'' planets to be those with $|\lambda|\leq20\degr$, while planets with $|\lambda|>20\degr$ are considered ``misaligned.'' We use this limit as a comparative diagnostic, but we acknowledge that the typical maximum misalignment in single-star warm Jupiter systems is not yet well-determined enough to distinguish between an upper limit of e.g. $|\lambda|=20\degr$ vs. $|\lambda|=25\degr$. Our results do not substantially change based on the exact limit that is selected.

%In reality, some of the ``aligned'' systems may also be more misaligned than is suggested by their 2D sky-projected \textbf{stellar} obliquities. While this 2D angle does not fully represent the absolute distribution of \textbf{stellar} obliquities, it is useful for relative studies across populations.

\subsubsection{Initial Trends}
\begin{figure*}
    \centering
    \includegraphics[width=0.98\textwidth]{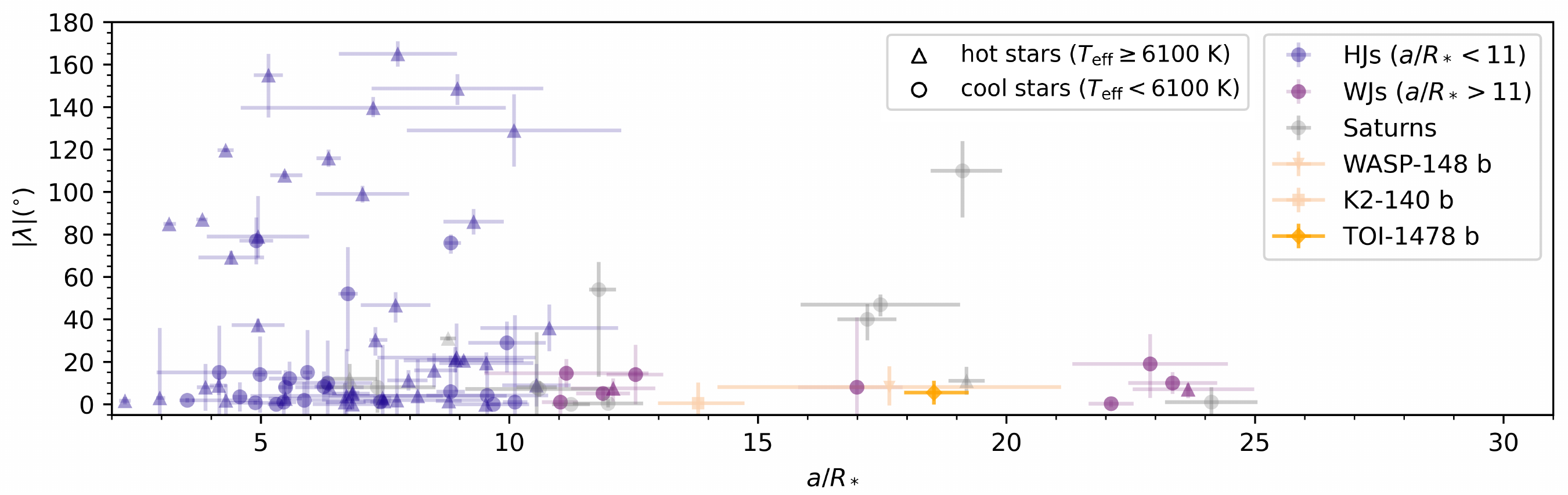}
    \caption{Distribution of measured $|\lambda|$ values for hot and warm Jupiters as a function of the orbital separation $a/R_*$. Measurements in systems below the Kraft break are shown as circular markers, while those above the Kraft break are displayed as triangular markers. The new measurement of TOI-1478 b from this work is shown in yellow, and the two previous SOLES measurements are shown in peach. Saturn-mass planets are shown in gray for reference; we note that WASP-148 b falls into this category, while K2-140 b and TOI-1478 b are more massive warm Jupiters.}
    \label{fig:WJ_vs_HJ}
\end{figure*}

Figure \ref{fig:WJ_vs_HJ} places the first three SOLES systems into context as some of the widest-separation tidally detached exoplanets with measured $|\lambda|$ values. All three of these systems are consistent with alignment. Furthermore, \textit{all} of the widest-orbiting ($a/R_*>11$), high-mass ($M\geq0.4M_J$) planets in single-star systems are, so far, consistent with alignment. We refer to this set of tidally-detached, high-mass planets as ``warm Jupiters'' throughout this work.

By contrast, many hot Jupiter systems ($a/R_*<11$, $M\geq0.4M_J$) are inconsistent with alignment. In particular, hot Jupiters orbiting hot stars (above the Kraft break at $T_{\mathrm{eff}}\approx6100$ K; \citep{kraft1967studies}) have been commonly discovered in misaligned configurations \citep[e.g.][]{albrecht2012obliquities}. Furthermore, recent work has demonstrated that, when their 3D stellar obliquities have been constrained, misaligned hot Jupiters are often found on polar orbits \citep{albrecht2021preponderance}. 

\begin{figure}
    \centering
    \includegraphics[width=0.48\textwidth]{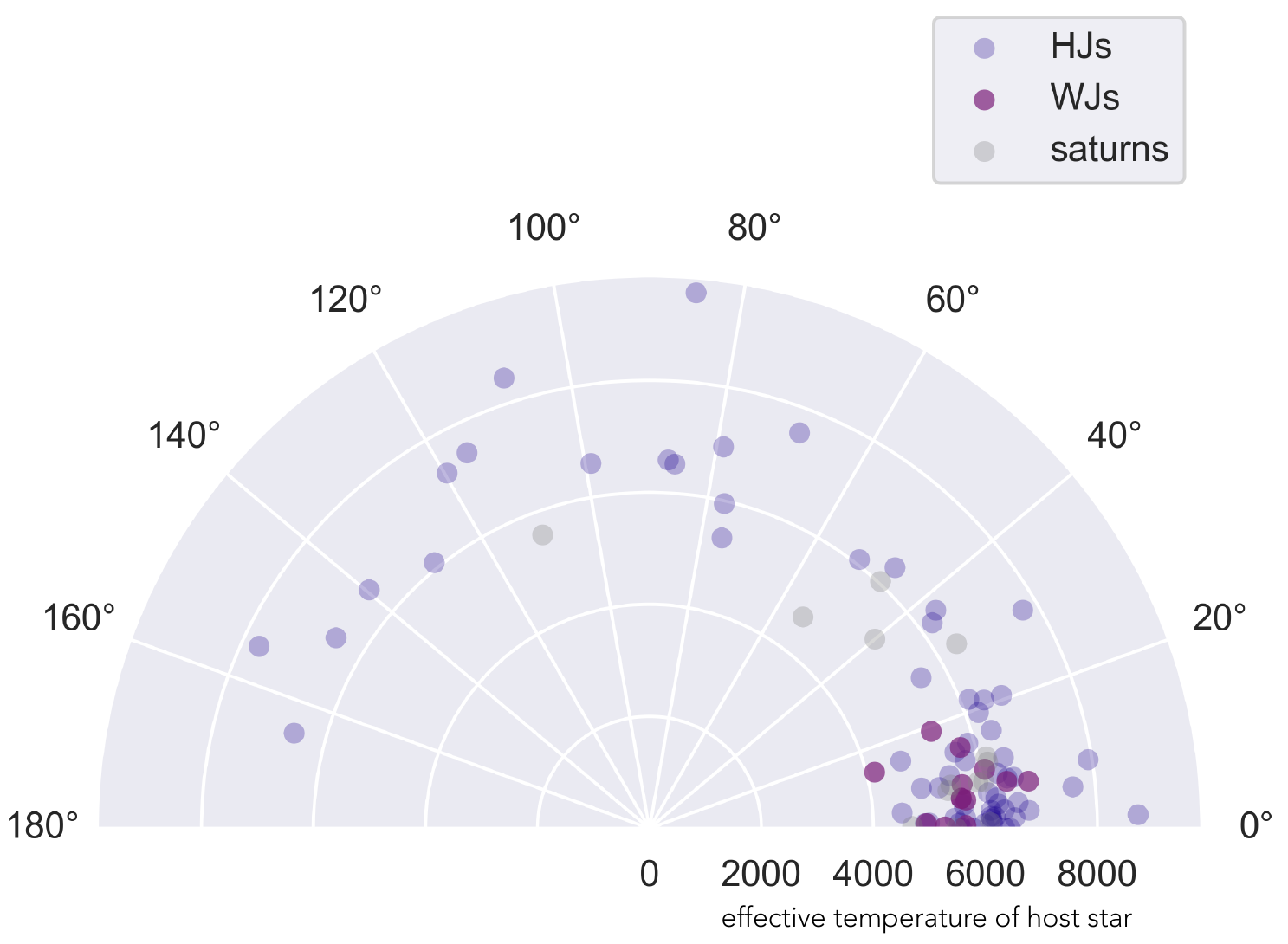}
    \caption{Distribution of sky-projected spin-orbit angles $|\lambda|$ observed for hot Jupiters (HJs), warm Jupiters (WJs), and Saturns. Unlike the hot Jupiters and Saturns, the warm Jupiters are confined to low values $|\lambda|\leq20\degree$ irrespective of their host star effective temperature.}
    \label{fig:polar_plot_distribution}
\end{figure}

This discrepancy points toward a distinct formation channel for hot Jupiters, which are likely dominated by a dynamically violent formation mechanism, as compared with warm Jupiters, which likely form quiescently (Wu, Rice, \& Wang, in review). The apparent distinction between these two populations is demonstrated in Figure \ref{fig:polar_plot_distribution}: hot Jupiters orbiting hot stars span a wide range of $|\lambda|$ values, while warm Jupiters have so far all been observed near alignment.

In parallel, we also find that lower-mass Saturns, with $0.2M_J\leq M_{\rm pl}<0.4M_J$, span a wider range of sky-projected spin-orbit angles than their higher-mass counterparts. While we turn our focus to the higher-mass systems for the remainder of our population study, we revisit the implications of this Saturn-mass population in Section \ref{subsection:extrasolar_saturns}.

\subsubsection{Statistical Significance: Divided Comparison}
\label{subsubsection:significance_binary}
Noting the visible distinction between hot and warm Jupiters, we tested the significance with which tidally detached Jupiters are more aligned than closer-in hot Jupiters.  To accomplish this, we compared the number of misaligned warm Jupiters with the number of misaligned hot Jupiters obtained from random draws. 

In total, our sample of warm Jupiters includes 12 systems, none of which are misaligned. For a direct comparison with this population, we iteratively drew 12 random $|\lambda|$ values without replacement from the full sample of hot Jupiters with measured $|\lambda|$ values. 

The sample $|\lambda|$ values were each allowed to vary within a Gaussian distribution characterized by the reported mean and uncertainty in $|\lambda|$ for that system. Asymmetric uncertainties were accounted for by first flipping a coin to determine whether we would sample from the upper or lower half of the distribution. We then sampled from a Gaussian distribution characterized by the corresponding uncertainty, and we calculated the distance of the sampled value from the mean. The final value was set to match that distance from the mean on the side of the distribution selected in the first step.

\begin{figure*}
    \centering
    \includegraphics[width=1.0\textwidth]{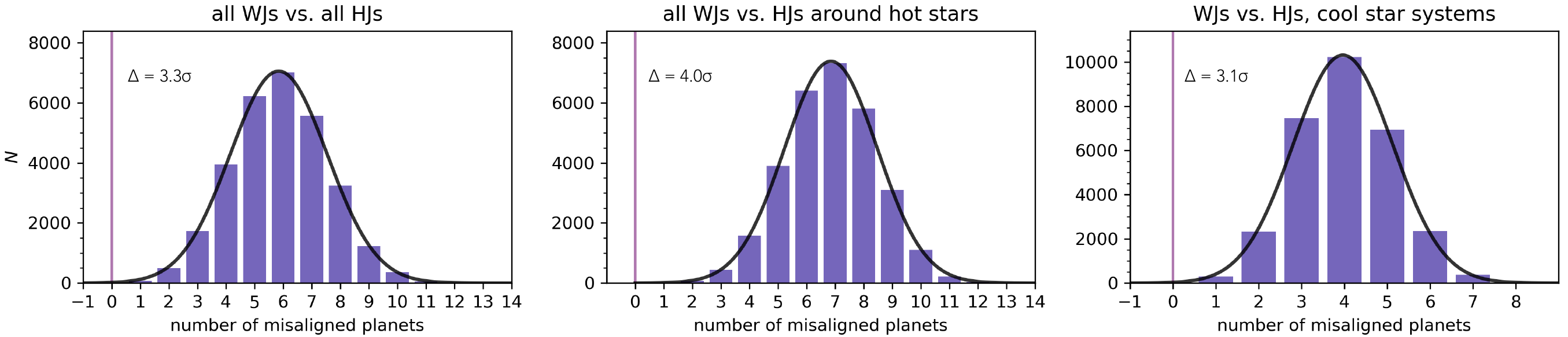}
    \caption{Significance with which warm Jupiters in single-star systems are more aligned than analogous hot Jupiters, considering three segmentations. Each panel shows the result of 30,000 iterations in which we sample $N_{\rm WJ}$ random $|\lambda|$ values without replacement from the set of hot Jupiter systems. The resulting distribution of misaligned hot Jupiter draws is shown in blue, in comparison with the number of misaligned warm Jupiter systems that have been observed (purple). The number of samples, $N_{\rm WJ}$, is determined by the number of observed warm Jupiters in each segmentation. \textit{Left:} The full population of warm Jupiters ($N_{\rm WJ}=12$) compared with the full population of hot Jupiters. \textit{Center:} The full population of warm Jupiters ($N_{\rm WJ}=12$) compared with the population of hot Jupiters orbiting hot stars. \textit{Right:} The population of warm and hot Jupiters orbiting cool stars ($N_{\rm WJ}=10$).}
    \label{fig:HJ_WJ_comparison}
\end{figure*}

We completed our comparative test with 30,000 iterations, then fit a Gaussian function to the resulting distribution of misaligned hot Jupiter draws. The result is shown in the left panel of Figure \ref{fig:HJ_WJ_comparison}.

This initial test demonstrates that, assuming no systematic differences between the two populations, warm Jupiters in single-star systems are more aligned than hot Jupiters in single-star systems to a significance of $\Delta=3.3\sigma$ (99.96\% of HJ trials included at least one misaligned planet). Therefore, warm Jupiters appear to be more aligned in these systems based on the current distribution of $|\lambda|$ measurements.

A shortcoming of this global population analysis, however, is that it does not consider the differential tidal effects in systems with different host star types. Short-period exoplanets orbiting stars with convective exteriors -- that is, cool stars below the Kraft break -- may be susceptible to tidal realignment through star-planet interactions. This effect should not extend to tidally detached warm Jupiter systems, since they orbit at a wide enough star-planet separation to evade strong tidal interactions.

In the tidal realignment framework, it would be more appropriate to compare the population of warm Jupiters with the population of hot Jupiters orbiting hot stars. Hot Jupiters orbiting stars above the Kraft break ($T_{\rm eff}\geq6100$ K) are less susceptible to tidal realignment and therefore may be more representative of the maximum stellar obliquity reached by each system. Repeating the same experiment but drawing hot Jupiters only from the hot star population ($T_{\mathrm{eff}}\geq6100$ K), we found that warm Jupiters in single-star systems are more aligned than hot Jupiters around single hot stars to $\Delta=4.0\sigma$ (100\% of HJ trials included at least one misaligned planet; see the central panel of Figure \ref{fig:HJ_WJ_comparison}). 

A final possibility is that the misalignment production mechanism may be unrelated to the dynamical evolution of planets within the system. Alternative mechanisms may act as a function of only the stellar properties: for example, if misalignments are caused by internal gravity waves within the host star \citep{rogers2012internal}, then the prevalence of misalignments should depend only on the host star's internal structure. In this case, misalignments should occur as a function of stellar type and should only be affected by the neighboring planet masses and orbital separations insofar as different stellar types produce different types of planets.

We examined this possibility by directly comparing the populations of hot and warm Jupiters around cool stars ($T_{\mathrm{eff}} \leq 6100$ K). To remove potential biases from tidal realignment within the hot Jupiter population, for this test we excluded systems with the largest ratio between their measured age and tidal realignment timescale. The tidal realignment timescale $\tau_R$ for cool stars, drawn from \citet{albrecht2012obliquities}, can be roughly estimated as

\begin{equation}
    \tau_{R} = 10\Big(\frac{a/R_*}{40}\Big)^6 \Big(\frac{M_{pl}}{M_{*}}\Big)^{-2} \,\,\mathrm{Gyr},
\end{equation}
where this expression has been scaled to match synchronized binary star systems with an assumed 10 Gyr main-sequence lifetime. 

We note that the tidal dissipation scaling for planets may substantially differ from that of stars. Furthermore, the ages within our sample have been heterogeneously measured, with large uncertainties in many cases. As a result, we conservatively excluded all systems from our sample with an estimated age greater than 10\% the tidal realignment timescale, $\tau_R$, of that system. This cut to the sample leaves 15 of the original 23 hot Jupiters around cool stars. 

The results of this final test are shown in the right panel of Figure \ref{fig:HJ_WJ_comparison}. Warm Jupiter systems with cool, single host stars are more aligned than analogous hot Jupiter systems with a significance of $\Delta=3.1\sigma$ (99.88\% of HJ trials included at least one misaligned planet). The difference between these two populations appears to be significant irrespective of the host star’s internal structure: that is, warm Jupiter systems are more aligned than hot Jupiter systems, even when considering only planets orbiting cool stars. This suggests that mechanisms that act primarily to misalign hot star systems, such as internal gravity waves \citep{rogers2012internal}, are likely not the sole driver of the observed misalignments.

However, further measurements are needed to verify the robustness of this final result. For example, Appendix C of \citet{albrecht2022stellar} describes potential caveats to the $|\lambda|$ measurements of the CoRoT-1 and CoRoT-19 systems. Excluding these, the significance of each result in Figure 4 is adjusted to $3.2\sigma$, $4.0\sigma$, and $2.4\sigma$ in the left, center, and right panel, respectively. While the first two results remain strong, the significance of the third result is weakened when these two systems are excluded.

``Misaligned’’ hot Jupiter systems around single, cool stars have also typically been interpreted as being misaligned due to the \textit{absence} of a large signal, rather than due to the presence of a large signal clearly indicating misalignment. This may signify a true, polar misalignment-- or, alternatively, it may instead indicate an underestimated $v\sin i_*$ value for the system, which can result from a measurement made using a spectrograph with insufficient resolving power (see Section \ref{subsection:comparison_stellar_parameters}). We discuss this point and further uncertainties within our analysis in greater detail in Section \ref{subsection:caveats}.

Because there are only two warm Jupiter systems with $|\lambda|$ measurements around hot stars, we are unable to repeat this experiment for hot star systems. Additional measurements are necessary to conclusively demonstrate whether the warm Jupiter population around hot stars substantially differs from its hot Jupiter counterpart.

%The weakened significance from this test may, however, result from the sparsity of hot stars hosting warm Jupiters in our sample (2 of the 12 systems), such that the comparison hot Jupiter sample is diluted by systems around cool stars that are commonly aligned in both the hot and warm Jupiter cases. Furthermore, internal gravity waves should lead to temporal variations in the $v\sin i_*$ measurements of stars, which have not yet been convincingly demonstrated \citep[e.g.][]{worku2022revisiting}. The prevalence of misaligned cool star systems among the wide-separation Saturn population (see Figure \ref{fig:WJ_vs_HJ}), as compared with the higher-mass warm Jupiters, also suggests that the misalignments originate through a dynamical process. While additional measurements are necessary to more thoroughly probe this possibility, we conclude that internal host star properties are likely not the primary driver of misalignments in our sample.

%By contrast, warm Jupiters should nearly always form aligned, except in the case of a tilted protoplanetary disk or violent dynamical interactions after disk dispersal. Our results demonstrate that the current populations of obliquities are fully consistent with this framework.

\subsubsection{Statistical Significance: Summed Comparison}
\label{subsubsection:significance_summed}
To further examine the robustness of our result, we also compared the summed $|\lambda|$ values between the single-star warm Jupiter population and the single-star hot Jupiter population. The benefit of this test is that it requires no specific assumptions regarding the cutoff between ``aligned'' and ``misaligned'' planets. Instead, we directly compare the summed $|\lambda|$ measurements to demonstrate population-wide differences.

\begin{figure*}
    \centering
    \includegraphics[width=1.0\textwidth]{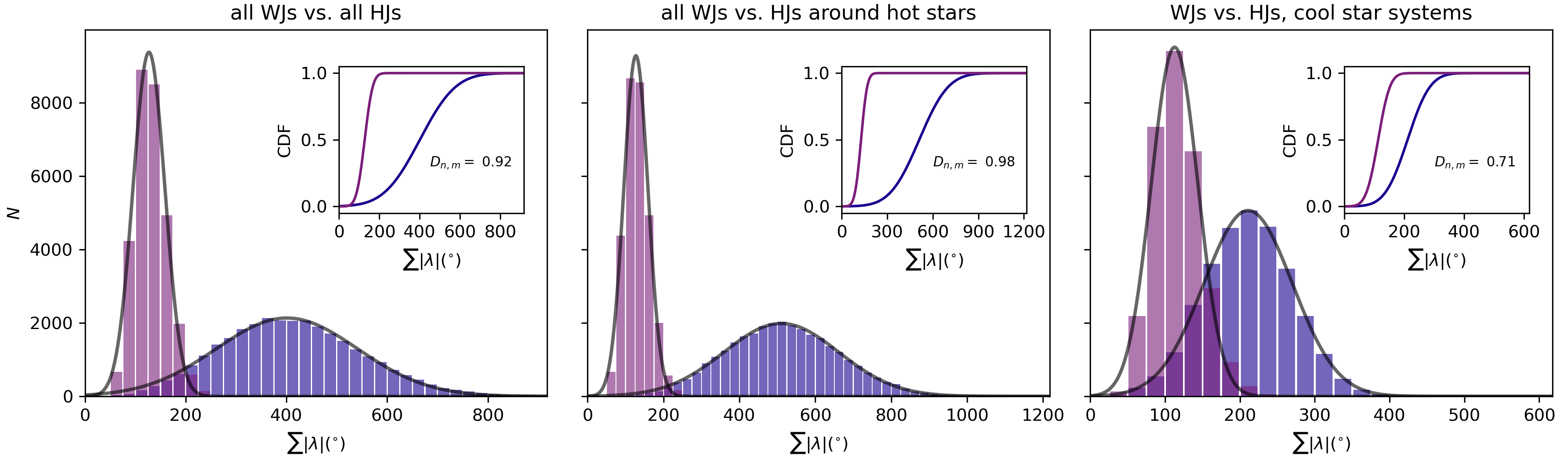}
    \caption{Comparison of the summed sky-projected misalignments, $\sum|\lambda|$, in warm Jupiter (purple) vs. hot Jupiter (blue) systems. The three segmentations shown are the same as those in Figure \ref{fig:HJ_WJ_comparison}; however, here we compare the distribution of summed sky-projected spin-orbit angles $\sum|\lambda|$, without assigning a label of ``aligned'' or ``misaligned'' to each system. Each distribution includes 30,000 draws, where each draw consists of $N_{\rm WJ}$ summed $|\lambda|$ measurements ($N_{\rm WJ}$=12 in the left and middle panels, and $N_{\rm WJ}$=10 in the right panel). Each sampled $|\lambda|$ value was allowed to vary about a Gaussian distribution characterized by its measured error bars.}
    \label{fig:HJ_WJ_comparison_summed}
\end{figure*}

We used the same three populations for comparison that were described in Section \ref{subsubsection:significance_binary}. However, in the summed tests, we added together the 12 randomly drawn $|\lambda|$ values from each population rather than assigning a label of ``aligned'' or ``misaligned.'' To account for uncertainties, we iteratively drew each $|\lambda|$ value from a Gaussian distribution characterized by the reported central value and its uncertainty (see Section \ref{subsubsection:significance_binary} for our treatment of asymmetric uncertainties). The resulting distributions are shown in Figure \ref{fig:HJ_WJ_comparison_summed}.

We fit a Gaussian function to each distribution and applied the two-sided Kolmogorov-Smirnov (K-S) test to demonstrate the likelihood that, in each panel of Figure \ref{fig:HJ_WJ_comparison_summed}, the warm and hot Jupiter systems were drawn from the same underlying distribution (the null hypothesis). The two-sided Kolmogorov-Smirnov statistic $D_{n,m}$ is given by 

\begin{equation}
    D_{n, m} = \mathrm{sup_x}|F_{1, n}(x) - F_{2, m}(x)|,
\end{equation}
where $F_{1, n}(x)$ and $F_{2, m}(x)$ are the two cumulative distributions with size $n$ and $m$, respectively, and $D_{n, m}$ is the supremum of the distances between these two distributions. The null hypothesis can be rejected with a significance characterized by $\alpha$ in the case that 
\begin{equation}
    D_{n, m} > \sqrt{-\frac{1}{2}\Big(\ln{\Big(\frac{\alpha}{2}\Big)}\Big)\Big(\frac{n+m}{nm}\Big)}.
\end{equation}
A smaller $\alpha$ value corresponds to a lower likelihood that the null hypothesis stands. That is, $\alpha$ quantifies the probability that the two sets of data would differ as much as they do if they were both randomly sampled from the same underlying distribution.

The cumulative distribution function (CDF) of a Gaussian is given by

\begin{equation}
    F(x) = \frac{1}{2}\Big[ 1 + \mathrm{erf}\Big(\frac{x - \mu}{\sigma\sqrt{2}}\Big)\Big],
\end{equation}
where $\mu$ and $\sigma$ are the mean and standard deviation, respectively, of the Gaussian. The CDF of each sample is shown in an inset on each panel, with the associated $D_{n,m}$ value provided to the right. 

We rule out the null hypothesis with a confidence level of $\alpha<10^{-16}$ in all cases, demonstrating that the two $|\lambda|$ distributions are distinct. In other words, warm Jupiter systems are preferentially more aligned, with a lower $\sum|\lambda|$, than hot Jupiter systems.

\vspace{5mm}
\subsection{Tidal Evolution of Stellar Obliquities}
Next, we directly examined the role of tidal realignments that may have altered the stellar obliquity distribution of single-star systems over time. One way to study this effect is by considering the $|\lambda|$ distribution as a function of the system age. If systems typically begin misaligned and are realigned by star-planet tidal interactions over time, as in the tidal realignment scenario, then older systems should, all else equal, be systematically more aligned than younger systems. 

\begin{figure}
    \centering
    \includegraphics[width=0.48\textwidth]{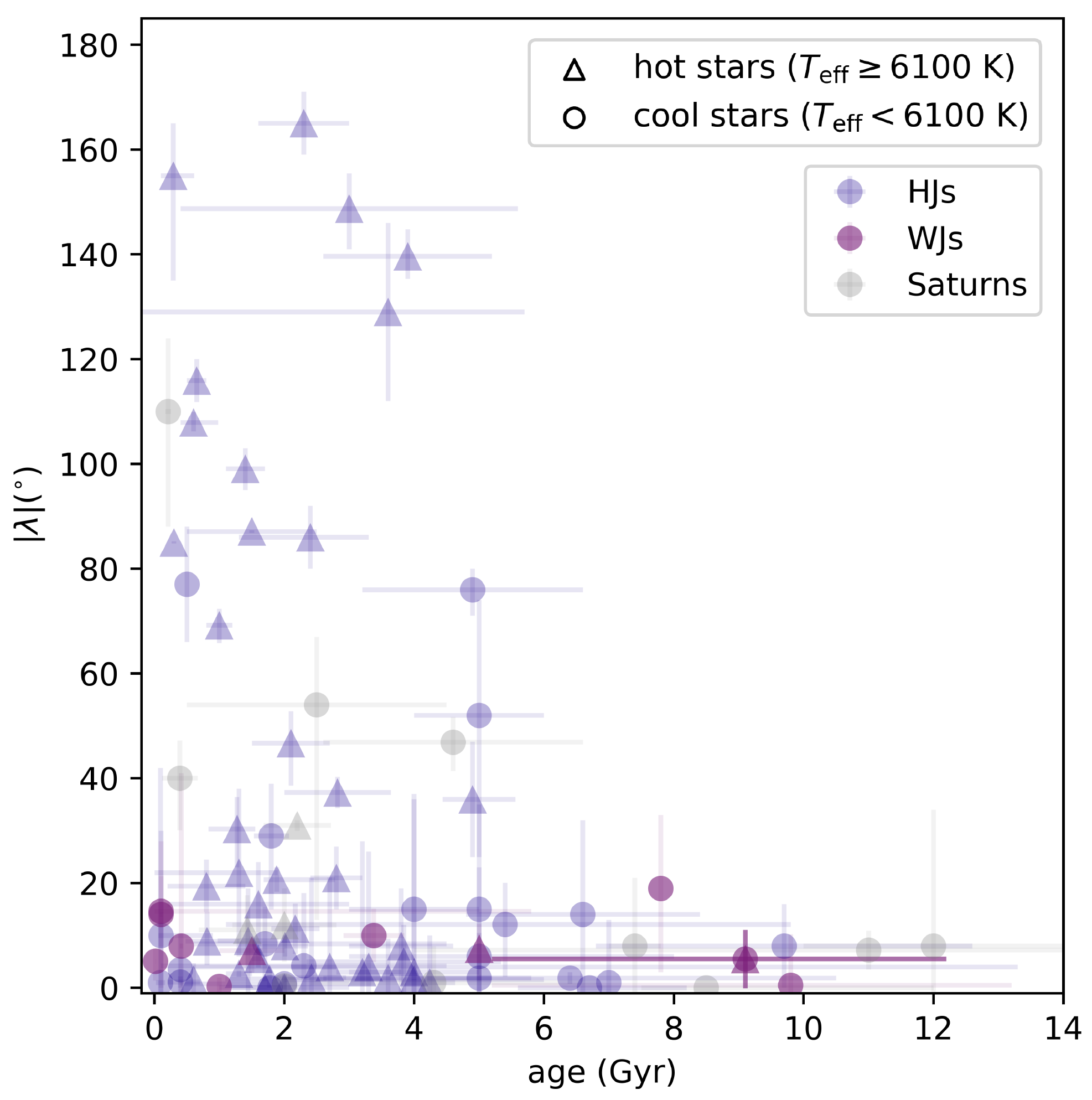}
    \caption{Sky-projected spin-orbit angles $|\lambda|$ as a function of the system age. Warm Jupiter systems (purple) are aligned irrespective of their age, consistent with a quiescent formation mechanism. By contrast, hot Jupiter systems (blue) are more commonly misaligned at young system ages. We emphasize that most observed misalignments are found in hot star systems, such that the apparent trend with age may instead indicate a trend with $T_{\rm eff}$.}
    \label{fig:age_v_lambda}
\end{figure}

The ages in our sample were, like other parameters described in Section \ref{subsection:sample_selection}, drawn from the NASA Exoplanet Archive and supplemented with values either directly from the literature or from the Extrasolar Values Encyclopaedia where possible. As shown in Figure \ref{fig:age_v_lambda}, hot Jupiters in single-star systems are typically aligned at ages $\tau>5$ Gyr, while they are sometimes misaligned at $\tau\leq5$ Gyr. By contrast, all warm Jupiters in our sample are aligned irrespective of their age, including several systems with $\tau\leq5$ Gyr.

The age distribution of hot Jupiter misalignments, paired with the absence of this trend in tidally detached warm Jupiter systems, is consistent with the framework of tidal realignment. However, degeneracies between stellar $T_{\rm eff}$ and age complicate the interpretation of Figure \ref{fig:age_v_lambda}. For example, cool stars are observed at a wider range of ages than hot stars due to their long main-sequence lifetimes -- reflected by the near-absence of hot stars with $\tau>5$ Gyr in our sample. Hot stars also cool over time after they finish burning hydrogen in their cores, such that initially hot stars may be observed with $T_{\rm eff}<6100$ K at later ages \citep{triaud2011time}. Therefore, the apparent trend with age observed among the hot Jupiter population may instead be more telling of intrinsic differences between hot and cool star systems \citep{safsten2020nature}.

\begin{figure*}
    \centering
    \includegraphics[width=0.98\textwidth]{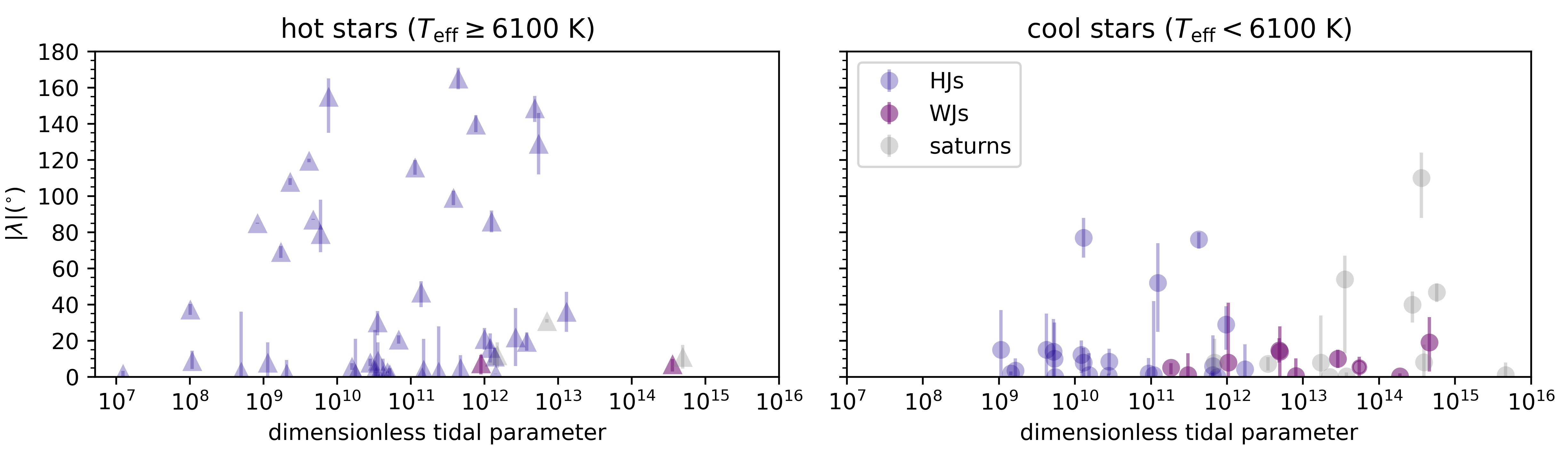}
    \caption{Measured $|\lambda|$ values as a function of the dimensionless tidal parameter $\xi = (M_{pl}/M_*)^{-2}(a/R_*)^6$, which is used as a proxy for the tidal realignment timescale. For clarity, we divide the full sample into hot star (triangles; $T_{\mathrm{eff}}\geq6100$ K) and cool star (circles; $T_{\mathrm{eff}}<6100$ K) systems; the same color scheme is applied to both panels. Warm Jupiters have comparatively long tidal realignment timescales, suggesting a primordial origin of their aligned orbits.}
    \label{fig:dimensionless_tidal_parameter}
\end{figure*}

To account for the heterogeneous stellar types within our sample, we next considered metrics for the tidal realignment timescale, characterized by the dimensionless tidal parameter $\xi = (M_{pl}/M_*)^{-2}(a/R_*)^6$. As shown in Figure \ref{fig:dimensionless_tidal_parameter}, hot Jupiters -- particularly those orbiting cool stars -- can have relatively short tidal realignment timescales, whereas warm Jupiters and Saturn-mass planets have comparatively longer tidal realignment timescales. 

In spite of their long tidal realignment timescales, warm Jupiters in single-star systems have not been found with large misalignments ($|\lambda|>20\degree$). Importantly, the scale factor for the planetary realignment timescale is not well-constrained in the equilibrium tides framework considered here. However, the young ages of many warm Jupiter systems (six with estimated ages $<2$ Gyr; see Figure \ref{fig:age_v_lambda}), paired with the prevalence of aligned warm Jupiters with relatively large $\xi$ -- despite the observation of some misaligned hot Jupiters with smaller $\xi$ values -- suggests that these systems were likely not all realigned over their lifetimes. Several systems containing Saturn-mass planets have been found with substantial misalignments, suggesting that these planets, which have some of the longest tidal realignment timescales in our sample, may be more susceptible to misalignment via dynamical interactions with neighboring planets.

\subsection{Caveats and Uncertainties}
\label{subsection:caveats}
We emphasize that our sample is heterogeneous: the measurements within our population study are drawn from many individual studies. Certain measurements would be particularly helpful to verify with additional observations: for example, our results are largely driven by the small sample of misaligned systems that have been observed, but the Rossiter-McLaughlin measurements of two misaligned hot Jupiter measurements in cool star systems include little to no egress data (e.g. CoRoT-19 b \citep{guenther2012transiting} and HATS-14 b \citep{zhou2015high}), providing weaker constraints on the obliquity of each system. Furthermore, underestimated $v\sin i_*$ values can mimic a low-amplitude Rossiter-McLaughlin signal, which may be incorrectly interpreted as a polar orbit. Additional tests to verify and more precisely constrain the obliquities of misaligned systems will be crucial to confirm or to challenge the results presented in this work.

Our cutoff between ``hot'' and ``warm'' Jupiters at $a/R_*>11$ was selected empirically and is not necessarily the ``true'' cutoff between the two populations. A shifted cutoff of $a/R_*>10$, which expands the sample of warm Jupiters by four systems (two with $|\lambda|>20\degree$), still reveals a tentative trend toward warm Jupiters being more aligned than hot Jupiters, with an adjusted significance of $2.7\sigma$, $3.7\sigma$, and $3.7\sigma$ for the left, center, and right panel of Figure 4, respectively. Different sets of stellar parameters may also shift individual systems slightly in $a/R_*$ space; for example, KELT-6 b is listed with $a/R_* = 11.17$ in \citet{albrecht2012obliquities}.

A transition range may exist between the hot and warm Jupiter populations, with a rate of misalignments that is reduced relative to that of the close-in hot Jupiters but that is higher than that of the tidally detached warm Jupiters. Additional $|\lambda|$ measurements for planets near $a/R_*\sim11$ will help to better sculpt the landscape of this parameter space and to potentially constrain the tidal realignment timescale for planets.

\section{Discussion} 
\subsection{Implications for Hot and Warm Jupiter Evolution}
We have demonstrated that, in single-star systems, the tidally detached warm Jupiter population is substantially more aligned than the hot Jupiter population, despite hot Jupiters' bias toward potentially realigned systems. This result has two key implications: (1) planets in single-star systems typically form within aligned protoplanetary disks, and (2) warm Jupiters form quiescently, whereas hot Jupiters are likely misaligned after the protoplanetary disk has dispersed.

%The current set of obliquity measurements is fully consistent with the framework of quiescent warm Jupiter formation. 
Quiescently formed warm Jupiters should remain roughly within the plane of their natal protoplanetary disk. As a result, if misalignments typically originate at the protoplanetary disk stage, then quiescently formed warm Jupiters should be observed with a similarly high rate of misalignments to that of the hot Jupiter population. In this scenario, observational biases should favor a \textit{higher} rate of misaligned warm Jupiters, since these tidally detached systems are unable to realign over timescales within the age of the Universe.

By contrast, the population of tidally detached warm Jupiters is significantly more aligned than the population of close-orbiting hot Jupiters. We conclude that warm Jupiters in single-star systems typically form within disks that are aligned with the host star's equator, whereas misalignments in hot Jupiter systems typically arise after the disk dispersal phase. This framework is consistent with the high rate of observed planetary companions in quiescent warm Jupiter systems \citep{huang2016warm}, by contrast with a much lower companion rate in hot Jupiter systems \citep{steffen2012kepler} that may have lost their companions during the misalignment process.

This framework is also supported by previous observations from the Atacama Large Millimeter Array (ALMA) that have suggested that most, if not all, protoplanetary disks are aligned with the spin axis of their host stars \citep{davies2019star}. Furthermore, no misaligned planets have been found in very young systems with $\tau<100$ Myr \citep{albrecht2022stellar}. While the sample of planets in this parameter space is small (only 6 to date), the absence of misaligned planets in young systems is consistent with our framework in which misalignments originate later in the systems' evolution. %A larger median galactic velocity dispersion has also been found for misaligned hot Jupiters as compared with aligned hot Jupiters \citep{hamer2022evidence}, providing further evidence for the late production of hot Jupiter misalignments.

%Our results demonstrate that the current set of obliquity measurements is fully consistent with the framework of quiescent warm Jupiter formation. Quiescently formed warm Jupiters should generally be aligned if they formed within initially aligned protoplanetary disks. 
Previous results have demonstrated that hot Jupiters' observed misalignments are fully consistent with a dominant formation mechanism that is dynamically violent \citep{rice2022origins}. Many hot Jupiters that are currently observed in aligned configurations may have been misaligned earlier in their evolution and subsequently realigned over time through star-planet tidal interactions. The closest-orbiting hot Jupiters around the coolest stars are typically all aligned, suggesting that tidal interactions have played an important role in the orbital evolution of these systems \citep{wang2021aligned}.

Some promising hot Jupiter misalignment mechanisms, such as the Kozai-Lidov mechanism in stellar binaries \citep{naoz2012formation}, require the presence of a stellar companion. However, the high rate of hot Jupiter misalignments, even in single-star systems, indicates that hot Jupiters cannot be misaligned exclusively through interactions with a stellar companion. 

Instead, planet-planet interactions likely play an important role in the production of hot Jupiters' misalignments. Alternative mechanisms invoking one or more planetary companions include secular chaos in multiplanet systems with overlapping resonances \citep{wu2011secular, teyssandier2019formation} and planet-planet scattering \citep{rasio1996dynamical}, particularly when combined with the Kozai-Lidov mechanism \citep{naoz2011hot, nagasawa2011orbital}. 

\subsection{The Evolution of Extrasolar Saturns}
\label{subsection:extrasolar_saturns}
While examining the hot and warm Jupiter populations, we also found that Saturn-mass planets, including those at a wide orbital separation $a/R_*>11$, span a wide range of spin-orbit angles. Three of the four misaligned Saturns in our sample (TOI-1268 b,\footnote{\citet{dong2022neid} includes 3 separate $|\lambda|$ measurements for TOI-1268 b. While we adopt the $|\lambda|=40^{+7.2}_{-9.9}\degree$ value included in the TEPcat catalogue, we note that the other two provided measurements each indicate a lower sky-projected obliquity ($|\lambda|=14^{+14}_{-10}\degree$ and $|\lambda|=25\pm13\degree$, from two separate reductions of the same NEID dataset). As a result, this system may not be strongly misaligned.} Kepler-63 b, and WASP-117 b) have significant projected eccentricities ($e>0.1$), suggesting that their eccentricities and stellar obliquities may have been jointly elevated. Three processes that can produce eccentric, misaligned Saturns include the adiabatic disk-dispersal resonance proposed by \citet{petrovich2020disk}, Kozai-Lidov interactions \citep{naoz2011hot}, and secular chaos \citep{wu2011secular}.

\subsubsection{Disk-Dispersal Resonance}
The \citet{petrovich2020disk} disk-dispersal resonance was originally proposed as a mechanism to naturally produce polar Neptunes—a growing population of observed systems \citep{stefansson2021warm, albrecht2022stellar}. This resonance occurs when the nodal precession rate of an outer planetary companion, induced by interactions with the protoplanetary disk, reaches commensurability with the nodal precession rate of the inner planet, induced by general relativistic effects and the stellar quadrupole field.

The same mechanism can also extend to higher-mass planets, with the caveat that a larger initial mutual inclination is required between the inner planet and its outer companion to ensure conservation of the angular momentum deficit $\Theta\prime$:

\begin{equation}
	\Theta\prime = L_{\rm in}(1 - \cos i_{\rm in}) + L_{\rm out}(1 - \cos i_{\rm out}).
\end{equation} 
Here, $L_{\rm in}$ and $L_{\rm out}$ are the orbital angular momenta of the inner and outer planet, while $i_{\rm in}$ and $i_{\rm out}$ are the planets' inclinations. 

The mutual inclination $i_{\rm mut}$ required to conserve $\Theta\prime$ is given by

\begin{equation}
i_{\rm mut} \geq \Big(\frac{m_{\rm in}}{m_{\rm out}}\Big)^{1/2}\Big(\frac{a_{\rm in}}{a_{\rm out}}\Big)^{1/4}.
\end{equation} 
Here, $m_{\rm in}$ and $m_{\rm out}$ are the masses of the inner and outer planet, respectively, while $a_{\rm in}$ and $a_{\rm out}$ are the corresponding semimajor axes. For a Neptune-mass planet at $a_{\rm in} = 0.1$ au with a $4M_J$ companion at $a_{\rm out}=5$ au, the required mutual inclination is only $i_{\rm mut}\geq2.5\degree$. By contrast, $i_{\rm mut}\geq5.9\degree$ if the inner planet is set to be Saturn-mass. 

Therefore, the population of misaligned Saturns may represent the upper mass limit of inner planets to which this mechanism applies. This would also imply a limit to the typical $i_{\rm mut}$ values attained during disk dispersal, with a larger $i_{\rm mut}$ limit required for lower-mass outer companions.

\subsubsection{Kozai-Lidov}
\label{subsubsection:kozai-lidov}
The Saturns in our sample may instead be misaligned by the Kozai-Lidov mechanism. In this framework, a third perturber with a significant initial mutual inclination ($i_{\rm mut}>39.2\degree$) is required to initialize a secular exchange of eccentricity and inclination between hierarchical orbits \citep{naoz2016eccentric}. In single-star systems, this additional perturber could be a companion planet within the system. However, this raises the question of why Saturn-mass planets would be misaligned, while warm Jupiters would not. 

If Saturns are misaligned through Kozai-Lidov oscillations, this would imply that Saturns more commonly appear in systems that meet the mechanism's required initial conditions. Planet-planet scattering simulations have offered evidence that, in systems with equal-mass planets, lower-mass planets preferentially reach higher mutual inclinations from planet-planet scattering. For example, \citet{raymond2010planet} found that 10\% of unstable planets in their disk-free simulations reached inclinations larger than 39$\degree$ in J-J-J systems (3 Jupiters), 42$\degree$ in S-S-S systems (3 Saturns), and 49$\degree$ in N-N-N systems (3 Neptunes). This increase in mutual inclinations with decreasing planet mass reflects the ease with which higher-mass planets can eject each other from the system, reducing the total number of close encounters over the systems' evolution. Therefore, the misaligned Saturns within our sample may represent a mass cutoff below which scattering processes can produce mutual inclinations $i_{\rm mut}>39.2\degree$.

%\textbf{Alternatively, the Saturns may originate in systems with unequal-mass planets. However, \citet{raymond2010planet} also demonstrated that these heterogeneous systems less efficiently produce large mutual inclinations.}

So far, no planetary companions have been found in any of the single-star Saturn systems that show evidence for misalignment. However, long-period Saturn-mass companions also cannot be ruled out based on current radial velocity surveys (e.g. Figure 1 in \citet{fulton2021california}). If the relatively high occurrence rate of misaligned Saturn-mass planets emerges due to planet-planet scattering combined with the Kozai-Lidov mechanism, we anticipate that outer planetary companions should be prevalent in misaligned Saturn systems.

%A wide range of spin-orbit angles for Saturn-mass planets may, therefore reflect a natural outcome of planet-planet scattering. However, this process alone cannot fully account for the misaligned Saturn population, since planet-planet scattering rarely produces misalignments greater than $\sim35\degree$.

\subsubsection{Secular Chaos}
Secular chaos may also play a role in misaligning the Saturns observed in our sample. Secular chaos requires the presence of at least $2-3$ planets in an individual system, and it is produced by overlapping, nonlinear secular resonances in which the precession rates of different planets reach commensurability \citep{lithwick2011theory}. Previous work has demonstrated that secular chaos can produce substantial spin-orbit misalignments in multiplanet systems \citep{wu2011secular}.

If Saturns form with neighboring companion planets, they may be misaligned through secular chaos in certain cases. However, spin-orbit angles larger than 90$\degree$ are difficult to generate from secular chaos \citep{wu2011secular, teyssandier2019formation}, such that highly misaligned systems (for example, Kepler-63, with $\lambda=-110^{+22}_{-14}\degree$ \citep{sanchis2013kepler}) likely could not have been misaligned by this mechanism. Furthermore, the secular chaos framework does not explain the heightened rate of spin-orbit misalignments for Saturn-mass planets as compared with warm Jupiters.

Because secular chaos takes place over timescales similar to or longer than the secular precession timescale, this framework would imply that misaligned Saturns should typically be observed at older stellar ages. Misaligned hot Jupiters have recently been found to be more common at older stellar ages \citep{hamer2022evidence}, providing evidence that secular chaos may be prevalent in at least some misaligned giant planet systems.

\section{Conclusions} 
In this work, we have measured the Rossiter-McLaughlin effect across the transit of tidally detached warm Jupiter TOI-1478 b using the Keck/HIRES and WIYN/NEID instruments. This is the third measurement made as part of the SOLES survey, which is designed to constrain planetary evolution models by studying stellar obliquities in long-period exoplanet systems. 

We combined this measurement with archival stellar obliquity constraints in single-star systems to conduct a population study demonstrating the statistical significance with which warm Jupiters are more aligned than analogous hot Jupiters. Our findings are summarized as follows:

\begin{itemize}
    \item The TOI-1478 system is consistent with alignment, with sky-projected spin-orbit angle $\lambda=6.2_{-5.5}^{+5.9}\degree$.
    \item The previous $v\sin i_*$ value for TOI-1478 may have been overestimated, as indicated by the low Rossiter-McLaughlin amplitude of the system.
    \item The current set of stellar obliquity constraints indicates that warm Jupiters in single-star systems are preferentially more aligned than analogous hot Jupiters (at $3.3\sigma$ when considering the full population, and at $4.0\sigma$ when comparing to hot stars that presumably have not experienced tidal realignment).
    \item Even when considering only systems around cool stars, warm Jupiters still appear to be more misaligned than analogous hot Jupiters (at $3.1\sigma$). However, further measurements are needed to better constrain the prevalence of misaligned hot Jupiters in cool, single-star systems and to verify the robustness of this result.
    \item Saturn-mass planets span a wide range of sky-projected spin-orbit angles, even at wide orbital separation.
\end{itemize}

Based on these findings, we conclude that (1) planets in single-star systems tend to form within aligned protoplanetary disks and (2) warm Jupiters tend to form quiescently, while hot Jupiters are misaligned after protoplanetary disk dispersal. We have also proposed several potential explanations for the preferential misalignment of Saturn-mass planets, including a secular disk-dispersal resonance, the Kozai-Lidov mechanism, or secular chaos.

Additional stellar obliquity measurements for warm Jupiters in binary star systems are needed to evaluate whether the observed trends extend beyond single-star systems. Further warm Jupiter spin-orbit angle measurements will also help to elucidate any potential correlations between eccentricities and stellar obliquities, which may provide improved insights into the key mechanisms contributing to extrasolar system misalignments.

\section{Acknowledgements}
\label{section:acknowledgements}

We thank our anonymous reviewer for their comments that have strengthened this manuscript. Over the duration of this project, M.R. was supported by the Heising-Simons Foundation 51 Pegasi b Fellowship and by the National Science Foundation Graduate Research Fellowship Program under Grant Number DGE-1752134. The data presented herein were obtained at the W. M. Keck Observatory, which is operated as a scientific partnership among the California Institute of Technology, the University of California and the National Aeronautics and Space Administration. The Observatory was made possible by the generous financial support of the W. M. Keck Foundation. This work is supported by the Astronomical Big Data Joint Research Center, co-founded by National Astronomical Observatories, Chinese Academy of Sciences and Alibaba Cloud. This research has made use of the Keck Observatory Archive (KOA), which is operated by the W. M. Keck Observatory and the NASA Exoplanet Science Institute (NExScI), under contract with the National Aeronautics and Space Administration. This research has made use of the NASA Exoplanet Archive, which is operated by the California Institute of Technology, under contract with the National Aeronautics and Space Administration under the Exoplanet Exploration Program. This paper contains data taken with the NEID instrument, which was funded by the NASA-NSF Exoplanet Observational Research (NN-EXPLORE) partnership and built by Pennsylvania State University. NEID is installed on the WIYN 3.5m telescope at KPNO, which is managed by the Association of Universities for Research in Astronomy (AURA) under a cooperative agreement with the NSF. The NEID archive is operated by the NASA Exoplanet Science Institute at the California Institute of Technology. We thank the NEID Queue Observers and WIYN Observing Associates for their skillful execution of our NEID observations. The authors are honored to be permitted to conduct astronomical research on Iolkam Du’ag (Kitt
Peak), a mountain with particular significance to the Tohono O’odham. 

\software{\texttt{numpy} \citep{oliphant2006guide, walt2011numpy, harris2020array}, \texttt{matplotlib} \citep{hunter2007matplotlib}, \texttt{pandas} \citep{mckinney2010data}, \texttt{scipy} \citep{virtanen2020scipy}, \texttt{allesfitter} \citep{gunther2020allesfitter}, \texttt{emcee} \citep{foremanmackey2013}, \texttt{MESA} \citep{paxton2010modules, paxton2013modules, paxton2015modules}}

\facility{Keck: I (HIRES), WIYN (NEID), Exoplanet Archive, Extrasolar Planets Encyclopaedia}

\bibliography{bibliography}
\bibliographystyle{aasjournal}

\startlongtable
\begin{deluxetable*}{lcccccccc}
\tablecaption{Parameters for the population of warm Jupiters, hot Jupiters, and Saturns studied in this work. Systems are ordered by $T_{\mathrm{eff}, *}$. Uncertainties in all parameters except $\lambda$ were used only for display purposes. Where uncertainties were not available, we used the central value for the system. Upper limits are provided from the literature where possible. \label{tab:population_parameters}}
\tabletypesize{\scriptsize}

\tablehead{
\colhead{} & \colhead{System} & \colhead{Planet} & \colhead{$M_{\rm pl}$ ($M_J$)} & \colhead{$a/R_*$} & \colhead{$e$} & \colhead{Age (Gyr)} & \colhead{$T_{\mathrm{eff}, *}$ (K)} & \colhead{$\lambda\, (\degr)$}}
\tablewidth{300pt}
\startdata
Warm Jupiters: & & & & & & & & \\
& WASP-80 & b & $0.538^{+0.035}_{-0.036}$ & $12.54^{+0.56}_{-0.58}$ & $0.002^{+0.010}_{-0.002}$ & $0.1^{+0.03}_{-0.02}$ & $4145\pm100$ & $-14\pm14$ \\
& WASP-53 & b & $2.132^{+0.092}_{-0.094}$ & $11.02^{+0.35}_{-0.37}$ & $<0.03$ & - & $4950\pm60$ & $-1\pm12$ \\
& WASP-84 & b & $0.694\pm0.028$ & $22.11\pm0.45$ & $0$ & $1$ & $5280\pm80$ & $-0.3\pm1.7$ \\
& HAT-P-17 & b & $0.58\pm0.06$ & $22.89\pm1.56$ & $0.35\pm0.01$ & 7.8 & $5322\pm55$ & $19^{+14}_{-16}$ \\
& TOI-1478 & b & $0.851^{+0.052}_{-0.047}$ & $18.54_{-0.6}^{+0.7}$ & $0.024^{+0.032}_{-0.017}$ & $9.1^{+3.1}_{-3.9}$ & $5597^{+83}_{-82}$ & $6.2_{-5.5}^{+5.9}$  \\
& HD 63433 & b & $<5$ & $16.99^{+0.92}_{-1.18}$ &$0$ & $0.414\pm0.023$ & $5640\pm74$ & $8^{+33}_{-45}$ \\
& K2-140 & b & $0.93\pm0.04$ & $13.79^{+0.93}_{-0.81}$ & $0$ & $9.8^{+3.4}_{-4.6}$ & $5654\pm55$ & $0.5\pm9.7$ \\
& HIP 67522 & b & $<5$ & $11.88\pm0.54$ & $0.059^{+0.193}_{-0.046}$ & $0.017\pm0.002$ & $5675\pm75$ & $5.1^{+2.5}_{-3.7}$ \\
& WASP-25 & b & $0.44\pm0.10$ & $11.14\pm1.65$ & $0$ & $0.1^{+5.7}_{-0.1}$ & $5736\pm35$ & $14.6\pm6.7$ \\
& HD 17156 & b & $3.51\pm0.21$ & $23.34\pm0.90$ & $0.68$ & $3.38\pm0.47$ & $6079\pm56$ & $10\pm5.1$ \\
& WASP-38 & b & $3.44\pm0.36$ & $12.09\pm0.85$ & $0.03$ & $>5.0$ & $6436\pm60$ & $7.5^{+4.7}_{-6.1}$ \\
& KOI-12 & b & $1.1^{+3.5}_{-0.8}$ & $23.65^{+1.33}_{-1.11}$ & $0.34^{0.08}_{-0.07}$ & $1.5\pm0.5$ & $6820\pm120$ & $-7.1^{+4.2}_{-2.8}$ \\
Hot Jupiters: & & & & & & & \\
& WASP-43 & b & $1.78\pm0.1$ & $4.58\pm0.15$ & $0$ & $0.4$ & $4520\pm120$ & $3.5\pm6.8$ \\
& Qatar-2 & b & $2.494\pm0.054$ & $5.935\pm0.094$ & $0$ & $5.0$ & $4645\pm50$ & $15\pm20$ \\
& Qatar-1 & b & $1.294^{+0.052}_{-0.049}$ & $6.27\pm0.19$ & $0$ & $1.7^{+2.8}_{-1.1}$ & $4910\pm100$ & $-8.4\pm7.1$ \\
& TOI-942 & b & $<2.6$ & $10.11^{+0.25}_{-0.24}$ & $0$ & $0.09\pm0.07$ & $4928^{+125}_{-85}$ & $1^{+41}_{-33}$ \\
& WASP-52 & b & $0.46\pm0.02$ & $7.40\pm0.20$ & $0$ & $0.4^{+0.3}_{-0.2}$ & $5000\pm100$ & $1.1\pm1.1$ \\
& HATS-2 & b & $1.35\pm0.15$ & $5.50\pm0.14$ & $0$ & $9.7\pm2.9$ & $5227\pm95$ & $8\pm8$ \\
& HATS-14 & b & $1.07\pm0.07$ & $8.82^{+0.20}_{-0.12}$ & $<0.142$ & $4.9\pm1.7$ & $5346\pm60$ & $76^{+4}_{-5}$ \\
& CoRoT-18 & b & $3.47\pm0.38$ & $6.34\pm0.89$ & $<0.08$ & $0.10^{+0.80}_{-0.04}$ & $5440\pm100$ & $-10\pm20$ \\
& WASP-19 & b & $1.154^{+0.078}_{-0.080}$ & $3.52^{+0.15}_{-0.16}$ & $0.0126^{+0.0140}_{-0.0089}$ & $6.4^{+4.1}_{-3.5}$ & $5460\pm90$ & $-1.9\pm1.1$ \\
& WASP-4 & b & $1.186^{+0.090}_{-0.098}$ & $5.46^{+0.25}_{-0.27}$ & $0$ & $7.0$ & $5540\pm55$ & $-1.0^{+14}_{-12}$ \\
& WASP-41 & b & $0.85\pm0.11$ & $9.95\pm0.78$ & $0$ & $1.80\pm0.27$ & $5546\pm33$ & $29^{+10}_{-14}$ \\
& WASP-47 & b & $1.142\pm0.023$ & $9.68\pm0.13$ & $0.0028\pm0.0028$ & $6.7^{+1.5}_{-1.1}$ & $5576\pm67$ & $0\pm24$ \\
& HAT-P-36 & b & $1.848\pm0.087$ & $4.98\pm0.11$ & $0.063\pm0.032$ & $6.6\pm1.8$ & $5620\pm40$ & $-14\pm18$ \\
& WASP-16 & b & $1.24\pm0.25$ & $9.55\pm1.23$ & $0$ & $2.3\pm2.2$ & $5630\pm70$ & $-4.2^{+11.0}_{-13.9}$ \\
& HAT-P-13 & b & $0.851\pm0.038$ & $5.872$ & $0.0133\pm0.0041$ & $5.0\pm0.8$ & $5653\pm90$ & $1.9\pm8.6$ \\
& WASP-5 & b & $1.58^{+0.13}_{-0.10}$ & $5.57\pm0.41$ & $0.038^{+0.026}_{-0.018}$ & $5.4^{+4.4}_{-4.3}$ & $5770\pm65$ & $12.1^{+8.0}_{-10.0}$ \\
& Kepler-17 & b & $2.45\pm0.11$ & $5.31\pm0.17$ & $<0.011$ & $<1.78$ & $5781\pm85$ & $0\pm15$ \\
& HAT-P-23 & b & $1.34\pm0.59$ & $4.16\pm1.26$ & $0.11\pm0.04$ & $4\pm1$ & $5885\pm72$ & $15\pm22$ \\
& CoRoT-1 & b & $1.03\pm0.12$ & $4.91\pm0.34$ & $0$ & $0.5$ & $5950\pm150$ & $77\pm11$ \\
& WASP-74 & b & $0.72\pm0.12$ & $4.89\pm0.53$ & $0$ & $2.0^{+1.0}_{-1.6}$ & $5984\pm57$ & $0.77\pm0.99$ \\
& WASP-28 & b & $0.907\pm0.043$ & $8.82\pm0.29$ & $0$ & $5^{+3}_{-2}$ & $6084\pm45$ & $6\pm17$ \\
& CoRoT-19 & b & $1.11\pm0.06$ & $6.75\pm0.19$ & $0.047\pm0.045$ & $5\pm1$ & $6090\pm70$ & $-52^{+27}_{-22}$ \\
& WASP-32 & b & $2.63\pm0.82$ & $7.74\pm1.73$ & $0$ & $2.42^{+0.53}_{-0.56}$ & $6100\pm100$ & $-2^{+17}_{-19}$ \\
& WASP-60 & b & $0.55\pm0.19$ & $10.09\pm2.16$ & $0$ & $3.6^{+2.1}_{-4.3}$ & $6105\pm50$ & $-129\pm17$ \\
& WASP-103 & b & $1.490\pm0.088$ & $2.96^{+0.11}_{-0.07}$ & $0$ & $4\pm1$ & $6110\pm160$ & $3\pm33$ \\
& HD 209458 & b & $0.73\pm0.04$ & $8.78\pm0.26$ & $0$ & $4\pm2$ & $6117\pm50$ & $1.58\pm0.08$ \\
& K2-34 & b & $1.78\pm0.13$ & $6.69^{+0.41}_{-0.38}$ & $0$ & $4.24^{+0.39}_{-0.44}$ & $6131\pm47$ & $-1^{+10}_{-9}$ \\
& WASP-22 & b & $0.67\pm0.19$ & $8.93\pm1.59$ & $0.02\pm0.01$ & $1.3^{+0.6}_{-1.3}$ & $6153\pm46$ & $22\pm16$ \\
& WASP-71 & b & $1.39\pm0.33$ & $4.30\pm0.68$ & $0$ & $3.6^{+1.6}_{-1.0}$ & $6180\pm52$ & $-1.9^{+7.1}_{-7.5}$ \\
& Kepler-8 & b & $0.59^{+0.13}_{-0.12}$ & $6.84^{+0.35}_{-0.41}$ & $0$ & $3.84\pm1.5$ & $6213\pm150$ & $5\pm7$ \\
& WASP-62 & b & $0.52\pm0.08$ & $9.53\pm0.95$ & $0$ & $0.8\pm0.6$ & $6230\pm80$ & $19.4^{+5.1}_{-4.9}$ \\
& KELT-6 & b & $0.52\pm0.12$ & $10.80\pm1.39$ & $0.22\pm0.11$ & $4.90^{+0.66}_{-0.46}$ & $6246\pm88$ & $-36\pm11$ \\
& WASP-61 & b & $2.68\pm0.84$ & $8.15\pm1.79$ & $0$ & $2.7^{+0.1}_{-0.6}$ & $6250\pm150$ & $4.0^{+17.1}_{-18.4}$ \\
& HAT-P-2 & b & $8.62\pm0.17$ & $10.54\pm0.69$ & $0.5172\pm0.0019$ & $1.44\pm0.47$ & $6290\pm60$ & $9\pm10$ \\
& CoRoT-11 & b & $2.33\pm0.34$ & $6.84\pm0.80$ & $0$ & $2\pm1$ & $6343\pm72$ & $0.1\pm2.6$ \\
& HAT-P-9 & b & $0.749^{+0.064}_{-0.063}$ & $8.48^{+0.46}_{-0.40}$ & $0.084^{+0.052}_{-0.047}$ & $1.6\pm1.4$ & $6350\pm150$ & $-16\pm8$ \\
& HATS-3 & b & $1.071\pm0.136$ & $7.45^{+0.17}_{-0.18}$ & $0$ & $3.2^{+0.6}_{-0.4}$ & $6351\pm76$ & $3\pm25$ \\
& XO-4 & b & $1.42\pm0.19$ & $7.71\pm0.70$ & $0$ & $2.1\pm0.6$ & $6397\pm70$ & $-46.7^{+8.1}_{-6.1}$ \\
& WASP-190 & b & $1.0\pm0.1$ & $8.91\pm0.57$ & $0$ & $2.8\pm0.4$ & $6400\pm100$ & $21\pm6$ \\
& XO-3 & b & $7.29\pm1.19$ & $4.94\pm0.53$ & $0.29$ & $2.82\pm0.82$ & $6429\pm75$ & $37.3\pm3.0$ \\
& HAT-P-34 & b & $3.33^{+0.21}_{-0.20}$ & $9.52^{+0.88}_{-0.63}$ & $0.432^{+0.029}_{-0.027}$ & $1.7^{+0.4}_{-0.5}$ & $6442\pm88$ & $0\pm14$ \\
& WASP-87 & b & $2.18\pm0.15$ & $3.89\pm0.17$ & $0$ & $3.8\pm0.8$ & $6450\pm120$ & $-8\pm11$ \\
& NGTS-2 & b & $0.74^{+0.13}_{-0.12}$ & $7.97^{+0.38}_{-0.42}$ & $0$ & $2.17\pm0.37$ & $6450\pm50$ & $-11.3\pm4.8$ \\
& WASP-7 & b & $0.96\pm0.13$ & $9.28\pm0.61$ & $0$ & $2.4^{+0.9}_{-1.1}$ & $6520\pm70$ & $86\pm6$ \\
& TOI-2109 & b & $5.02\pm0.75$ & $2.27\pm0.11$ & $0$ & $1.77^{+0.88}_{-0.68}$ & $6530^{+160}_{-150}$ & $1.7\pm1.7$ \\
& WASP-17 & b & $0.78\pm0.23$ & $8.95\pm1.73$ & $0$ & $3.0\pm2.6$ & $6550\pm100$ & $-148.7^{+7.7}_{-6.7}$ \\
& HAT-P-56 & b & $2.31\pm0.44$ & $6.38\pm0.69$ & $<0.246$ & $2.01\pm0.35$ & $6566\pm50$ & $8\pm2$ \\
& HAT-P-6 & b & $1.32\pm0.3$ & $7.76\pm1.19$ & $0$ & $2.3\pm0.7$ & $6570\pm80$ & $165\pm6$ \\
& WASP-15 & b & $0.54\pm0.29$ & $7.26\pm2.67$ & $0$ & $3.9\pm1.3$ & $6573\pm70$ & $-139.6^{+4.3}_{-5.2}$ \\
& WASP-121 & b & $1.16\pm0.07$ & $3.82^{+0.10}_{-0.12}$ & $0$ & $1.5\pm1.0$ & $6586\pm59$ & $87.08^{+0.29}_{-0.27}$ \\
& WASP-79 & b & $0.85\pm0.18$ & $7.04\pm0.94$ & $0$ & $1.4\pm0.3$ & $6600\pm100$ & $-99.1^{+4.1}_{-3.9}$ \\
& WASP-66 & b & $2.35\pm0.98$ & $6.72\pm1.93$ & $0$ & $3.3^{+10.0}_{-2.7}$ & $6600\pm150$ & $-4\pm22$ \\
& XO-6 & b & $<4.4$ & $9.08\pm1.21$ & $0$ & $1.88^{+0.90}_{-0.20}$ & $6720\pm100$ & $-20.7\pm2.3$ \\
& KELT-7 & b & $1.39\pm0.22$ & $5.49\pm0.34$ & $0$ & $1.3\pm0.2$ & $6789^{+50}_{-49}$ & $2.7\pm0.6$ \\
& WASP-100 & b & $1.26\pm0.45$ & $4.94\pm1.03$ & $0$ & - & $6940\pm120$ & $79^{+19}_{-10}$ \\
& TOI-1518 & b & $<2.3$ & $4.29\pm0.16$ & $<0.01$ & - & $7300\pm100$ & $-119.66^{+0.93}_{-0.98}$ \\
& KELT-17 & b & $1.31^{+0.28}_{-0.29}$ & $6.36^{+0.25}_{-0.24}$ & - & $0.65\pm0.15$ & $7454\pm49$ & $-115.9\pm4.1$ \\
& MASCARA-1 & b & $3.7\pm0.9$ & $4.40\pm0.66$ & $0$ & $1.0\pm0.2$ & $7554\pm150$ & $69.2^{+3.1}_{-3.4}$ \\
& KELT-21 & b & $<3.91$ & $6.85\pm0.13$ & $0$ & $1.6\pm0.1$ & $7598^{+81}_{-84}$ & $-5.6^{+1.7}_{-1.9}$ \\
& TOI-1431 & b & $3.12\pm0.18$ & $5.15\pm0.29$ & $0.0022^{+0.0030}_{-0.0016}$ & $0.29^{+0.32}_{-0.19}$ & $7690^{+400}_{-250}$ & $-155^{+20}_{-10}$ \\
& HAT-P-69 & b & $3.58\pm0.58$ & $7.30^{+0.24}_{-0.12}$ & $0$ & $1.27^{+0.28}_{-0.44}$ & $7724^{+250}_{-360}$ & $30.3^{+6.1}_{-7.3}$ \\
& HATS-70 & b & $12.9^{+1.8}_{-1.6}$ & $4.15^{+0.16}_{-0.18}$ & $<0.18$ & $0.81^{+0.50}_{-0.33}$ & $7930^{+630}_{-820}$ & $8.9^{+5.6}_{-4.5}$ \\
& HAT-P-70 & b & $<6.78$ & $5.48^{+0.36}_{-0.29}$ & $0$ & $0.60^{+0.38}_{-0.20}$ & $8450^{+540}_{-690}$ & $107.9^{+2.0}_{-1.7}$ \\
& KELT-20 & b & $<3.382$ & $7.47^{+0.35}_{-0.41}$ & - & $0.6$ & $8730^{+250}_{-260}$ & $1.6\pm3.1$ \\
& KELT-9 & b & $2.88\pm0.84$ & $3.15^{+0.14}_{-0.12}$ & $0$ & $0.3$ & $9600\pm400$ & $-85.01\pm0.23$ \\
Saturns: & & & & & & & \\
& HAT-P-12 & b & $0.211\pm0.012$ & $11.80^{+0.35}_{-0.19}$ & 0 & $2.5\pm2.0$ & $4665\pm45$ & $-54^{+41}_{-13}$ \\
& WASP-69 & b & $0.29\pm0.03$ & $11.98\pm0.71$ & 0 & $2.0$ & $4700\pm50$ & $0.4^{+2.0}_{-1.9}$ \\
& TOI-1268 & b & $0.29$ & $17.20^{+0.58}_{-0.60}$ & $0.12^{+0.24}_{-0.12}$ & $0.39^{+0.28}_{-0.27}$ & $5257\pm40$ & $40^{+7.2}_{-9.9}$ \\
& WASP-6 & b & $0.37\pm0.08$ & $10.60\pm1.49$ & $0.05\pm0.02$ & $11\pm7$ & $5375\pm65$ & $7.2\pm3.7$ \\
& WASP-148 & b & $0.291\pm0.025$ & $17.64\pm3.46$ & $0.202\pm0.063$ & - & $5437\pm21$ & $-8.2^{+8.7}_{-9.7}$ \\
& WASP-39 & b & $0.28\pm0.03$ & $11.24\pm0.38$ & 0 & $8.5^{+3.5}_{-1.0}$ & $5485\pm50$ & $0\pm11$ \\
& Kepler-63 & b & $0.378$ & $19.11^{+0.80}_{-0.64}$  & 0.45 & $0.210\pm0.045$ & $5576\pm50$ & $-110^{+22}_{-14}$ \\
& WASP-21 & b & $0.300\pm0.011$ & $10.55\pm0.41$ & 0 & $12\pm2$ & $5924\pm55$ & $8^{+26}_{-27}$ \\
& WASP-13 & b & $0.36\pm0.09$ & $7.34\pm1.17$ & 0 & $7.4\pm0.4$ & $6025\pm21$ & $8^{+13}_{-12}$ \\
& WASP-117 & b & $0.30\pm0.05$ & $17.46\pm1.60$ & $0.30\pm0.02$ & $4.6\pm2.0$ & $6040\pm90$ & $-46.9^{+5.5}_{-4.8}$ \\
& HD 332231 & b & $0.244\pm0.021$ & $24.12^{+0.93}_{-0.94}$ & $0.032^{+0.030}_{-0.022}$ & $4.3^{+2.5}_{-1.9}$ & $6089^{+97}_{-96}$ & $-1\pm7$ \\
& HD 149026 & b & $0.38\pm0.06$ & $6.79\pm0.55$ & 0 & $2.0\pm0.8$ & $6147\pm50$ & $12\pm7$ \\
& K2-232 & b & $0.398\pm0.037$ & $19.20\pm0.36$ & $0.258\pm0.025$ & $1.43^{+0.82}_{-0.75}$ & $6154\pm60$ & $-11.1\pm6.6$ \\
& WASP-174 & b & $0.330\pm0.091$ & $8.77\pm0.16$ & 0 & $2.20\pm0.52$ & $6400\pm100$ & $31\pm1$ \\
\enddata
\end{deluxetable*}

\end{document}